\documentclass[10pt, letterpaper, oneside]{article}

\usepackage{amsmath}	
\usepackage{amstext}

\usepackage{amssymb}	
\usepackage{dsfont}     
\usepackage{mathrsfs}   
\usepackage{bm}         

\usepackage{amsthm}
\usepackage{thmtools, thm-restate}    
\usepackage{IEEEtrantools}

\usepackage{mathtools}  

\usepackage[inline]{enumitem}  

\usepackage{nameref}
\usepackage{hyperref}
\hypersetup{
    unicode=false,           
    pdftoolbar=false,         
    pdfmenubar=false,         
    pdffitwindow=false,      
    pdfstartview={FitH},     
    pdftitle={My title},     
    pdfauthor={Author},      
    pdfsubject={Subject},    
    pdfcreator={Creator},    
    pdfproducer={Producer},  
    pdfkeywords={keyword1} {key2} {key3}, 
    pdfnewwindow=true,       
    colorlinks=true,         
    linkcolor=BrickRed,      
    citecolor=blue,          
    filecolor=magenta,       
    urlcolor=cyan            
}
\usepackage{cleveref}

\usepackage[usenames,dvipsnames,svgnames,table]{xcolor}

\usepackage{graphicx}
\usepackage{caption}
\usepackage{subcaption}
\usepackage[export]{adjustbox}

\usepackage{accents}	

\usepackage[style=numeric, backend=bibtex, giveninits=true]{biblatex}
\defbibheading{bibliography}[\refname]{\section*{#1}}
\renewbibmacro{in:}{}

\usepackage[runin]{abstract}
\definecolor{lightSolarizedBase2}{RGB}{238,232,213}
\definecolor{lightSolarizedBase3}{RGB}{253,246,227}
\definecolor{lightSolarizedBase2a}{RGB}{245,242,231}
\definecolor{DimmedMolokaiBG}{RGB}{31,31,31}
\definecolor{WarmGrayBG}{RGB}{245,244,241}
\definecolor{DimmedMolokaiFG}{RGB}{185,188,186}
\definecolor{green}{RGB}{105,130,91}

\usepackage{setspace} 	

\usepackage{geometry}                   
\usepackage[pagestyles, explicit]{titlesec}

    \titleformat{\section}
                [hang]                              
                {\normalfont\scshape\filcenter}     
                {\thesection.}                      
                {5pt}                               
                {#1}                                
                []                                  
    \titlespacing{\section}
              {0pt} 
              {10pt} 
              {10pt} 
              [0pt] 

    \titleformat{\subsection} 
                [runin]                             
                {\normalfont}                       
                {\thesubsection}                    
                {5pt}                               
                {\bfseries #1}                      
                []                                  

    \titleformat{\paragraph}
                [runin]                             
                {\normalfont}                       
		{\thesubsection}
                {0pt}                               
                {}                                  
                []                                  
    \titlespacing{\paragraph}
              {0pt} 
              {0pt} 
              {0pt} 
              [0pt] 

    \newpagestyle{titlepage}[]{
        \sethead[]
                []
                []
                {}
                {}
                {}
        \setfoot[]
                []
                []
                {}
                {\tiny \thepage}
                {}
      }

    \newpagestyle{articleStyle}[]{
        \sethead[]
                []
                []
                {}
                {}
                {}
        \setfoot[]
                []
                []
                {}
                {\tiny \thepage}
                {}
      }

\declaretheoremstyle[
spaceabove=6pt, spacebelow=6pt,
headindent=0pt,
headfont=\normalfont\small\scshape,
notefont=\normalfont\bfseries, notebraces={}{},
bodyfont=\itshape,  
postheadspace=4pt,
headpunct={.},
qed=
]{THMsc}

\declaretheorem[heading=Theorem,			   style=DEFrm]{thm}
\declaretheorem[heading=Theorem,     numberlike=thm,   	   style=THMsc]{thmSc}
\declaretheorem[heading=Lemma,	     numberlike=thm,	   style=DEFrm]{lemma}
\declaretheorem[heading=Proposition, numberlike=thm, 	   style=THMsc]{proposition}

\declaretheoremstyle[
spaceabove=6pt, spacebelow=6pt,
headindent=0pt,
headfont=\normalfont\bfseries,
notefont=\normalfont, notebraces={}{},
bodyfont=\normalfont,   
postheadspace=1em,
headpunct={.},
qed=
]{DEFrm}
\declaretheorem[heading=Definition,  numberlike=thm, 	  style=DEFrm]{definition}
\declaretheorem[heading=Definition,  numberlike=thm, 	  style=DEFrm]{defn}
\declaretheorem[heading=Remark,	     numberlike=thm,      style=DEFrm]{remark}

\declaretheorem[heading=Assumption,  numberlike=thm, 	  style=DEFrm]{assumptionRm}
\declaretheorem[heading=Corollary,   numberlike=thm, 	  style=DEFrm]{corollaryRm}

\declaretheoremstyle[
spaceabove=6pt, spacebelow=6pt,
headindent=0pt,
headfont=\normalfont\bfseries,
notefont=\normalfont,
notebraces={(}{)},
bodyfont=\normalfont,
postheadspace=1em,
headpunct={.},
qed=
]{exrcs}

\declaretheoremstyle[
spaceabove=6pt, spacebelow=6pt,
headindent=0pt,
headfont=\normalfont,
notefont=\normalfont,
notebraces={(}{)},
bodyfont=\normalfont,
postheadspace=1em,
headpunct={.},
qed=
]{example}
\declaretheorem[heading=Example,                 style=example]{example}

\declaretheoremstyle[
spaceabove=2pt, spacebelow=12pt,
headindent=0pt,
headfont=\itshape,
notefont=\itshape, 
notebraces={}{},
bodyfont=\normalfont,
postheadspace=1em,
headpunct={.},
qed=  $\square$ 
]{prfs}
\declaretheorem[numbered=no, heading=Proof, style=prfs]{pf}

\declaretheoremstyle[
spaceabove=4pt, spacebelow=-6pt,
headindent=0pt,
headfont=\normalfont\small\scshape,
notefont=\normalfont\bfseries, notebraces={}{},
bodyfont=\itshape,  
postheadspace=4pt,
headpunct={.},
qed=  
]{SmallCapsShortShort}

\DeclarePairedDelimiter\abs{\lvert}{\rvert}               

\DeclarePairedDelimiter\norm{\lVert}{\rVert}               

\DeclarePairedDelimiterX\brk[2]{\langle}{\rangle}{#1,#2} 
\DeclarePairedDelimiterX\sbrk[1]{\langle}{\rangle}{#1} 

\DeclarePairedDelimiter\set{\{}{\}} 
\DeclarePairedDelimiterX\Set[2]{\{}{\}}{#1 \;;\; #2} 

\usepackage{xpatch}
\xapptocmd\normalsize{%
 \abovedisplayskip=6pt plus 3pt minus 4pt
 \abovedisplayshortskip=0pt plus 3pt
 \belowdisplayskip=6pt plus 3pt minus 4pt
 \belowdisplayshortskip=3pt plus 3pt minus 2pt
}{}{}

\usepackage{listings}

\definecolor{dkgreen}{rgb}{0,0.6,0}
\definecolor{gray}{rgb}{0.5,0.5,0.5}
\definecolor{mauve}{rgb}{0.58,0,0.82}

\DeclareMathSymbol{\widehatsym}{\mathord}{largesymbols}{"62}
\newcommand\lowerwidehatsym{%
  \text{\smash{\raisebox{-1.3ex}{%
    $\widehatsym$}}}}
\newcommand\wh[1]{%
  \mathchoice
    {\accentset{\displaystyle\lowerwidehatsym}{#1}}
    {\accentset{\textstyle\lowerwidehatsym}{#1}}
    {\accentset{\scriptstyle\lowerwidehatsym}{#1}}
    {\accentset{\scriptscriptstyle\lowerwidehatsym}{#1}}
}
\makeatletter
\DeclareRobustCommand\widecheck[1]{{\mathpalette\@widecheck{#1}}}
\def\@widecheck#1#2{%
    \setbox\z@\hbox{\m@th$#1#2$}%
    \setbox\tw@\hbox{\m@th$#1%
       \widehat{%
          \vrule\@width\z@\@height\ht\z@
          \vrule\@height\z@\@width\wd\z@}$}%
    \dp\tw@-\ht\z@
    \@tempdima\ht\z@ \advance\@tempdima2\ht\tw@ \divide\@tempdima\thr@@
    \setbox\tw@\hbox{%
       \raise\@tempdima\hbox{\scalebox{1}[-1]{\lower\@tempdima\box
\tw@}}}%
    {\ooalign{\box\tw@ \cr \box\z@}}}
\makeatother
\makeatletter
\DeclareRobustCommand\wc[1]{{\mathpalette\@widecheck{#1}}}
\def\@widecheck#1#2{%
    \setbox\z@\hbox{\m@th$#1#2$}%
    \setbox\tw@\hbox{\m@th$#1%
       \widehat{%
          \vrule\@width\z@\@height\ht\z@
          \vrule\@height\z@\@width\wd\z@}$}%
    \dp\tw@-\ht\z@
    \@tempdima\ht\z@ \advance\@tempdima2\ht\tw@ \divide\@tempdima\thr@@
    \setbox\tw@\hbox{%
       \raise\@tempdima\hbox{\scalebox{1}[-1]{\lower\@tempdima\box
\tw@}}}%
    {\ooalign{\box\tw@ \cr \box\z@}}}
\makeatother
\DeclareMathSymbol{\widetildesym}{\mathord}{largesymbols}{"65}
\newcommand\lowerwidetildesym{%
  \text{\smash{\raisebox{-1.3ex}{%
    $\widetildesym$}}}}
\newcommand\wt[1]{%
  \mathchoice
    {\accentset{\displaystyle\lowerwidetildesym}{#1}}
    {\accentset{\textstyle\lowerwidetildesym}{#1}}
    {\accentset{\scriptstyle\lowerwidetildesym}{#1}}
    {\accentset{\scriptscriptstyle\lowerwidetildesym}{#1}}
}
\makeatletter
\newcommand*\if@single[3]{%
  \setbox0\hbox{${\mathaccent"0362{#1}}^H$}%
  \setbox2\hbox{${\mathaccent"0362{\kern0pt#1}}^H$}%
  \ifdim\ht0=\ht2 #3\else #2\fi
  }
\newcommand*\rel@kern[1]{\kern#1\dimexpr\macc@kerna}
\newcommand*\widebar[1]{\@ifnextchar^{{\wide@bar{#1}{0}}}{\wide@bar{#1}{1}}}
\newcommand*\wide@bar[2]{\if@single{#1}{\wide@bar@{#1}{#2}{1}}{\wide@bar@{#1}{#2}{2}}}
\newcommand*\wide@bar@[3]{%
  \begingroup
  \def\mathaccent##1##2{%
    \if#32 \let\macc@nucleus\first@char \fi
    \setbox\z@\hbox{$\macc@style{\macc@nucleus}_{}$}%
    \setbox\tw@\hbox{$\macc@style{\macc@nucleus}{}_{}$}%
    \dimen@\wd\tw@
    \advance\dimen@-\wd\z@
    \divide\dimen@ 3
    \@tempdima\wd\tw@
    \advance\@tempdima-\scriptspace
    \divide\@tempdima 10
    \advance\dimen@-\@tempdima
    \ifdim\dimen@>\z@ \dimen@0pt\fi
    \rel@kern{0.6}\kern-\dimen@
    \if#31
      \overline{\rel@kern{-0.6}\kern\dimen@\macc@nucleus\rel@kern{0.4}\kern\dimen@}%
      \advance\dimen@0.4\dimexpr\macc@kerna
      \let\final@kern#2%
      \ifdim\dimen@<\z@ \let\final@kern1\fi
      \if\final@kern1 \kern-\dimen@\fi
    \else
      \overline{\rel@kern{-0.6}\kern\dimen@#1}%
    \fi
  }%
  \macc@depth\@ne
  \let\math@bgroup\@empty \let\math@egroup\macc@set@skewchar
  \mathsurround\z@ \frozen@everymath{\mathgroup\macc@group\relax}%
  \macc@set@skewchar\relax
  \let\mathaccentV\macc@nested@a
  \if#31
    \macc@nested@a\relax111{#1}%
  \else
    \def\gobble@till@marker##1\endmarker{}%
    \futurelet\first@char\gobble@till@marker#1\endmarker
    \ifcat\noexpand\first@char A\else
      \def\first@char{}%
    \fi
    \macc@nested@a\relax111{\first@char}%
  \fi
  \endgroup
}

\newdimen\CdotAxis
\newcommand*{\CdotAux}[3]{%
  {%
    \settoheight\CdotAxis{$#2\vcenter{}$}%
    \sbox0{%
      \raisebox\CdotAxis{%
        \scalebox{#1}{%
          \raisebox{-\CdotAxis}{%
            $\mathsurround=0pt #2#3$%
          }%
        }%
      }%
    }%
    \dp0=0pt %
    \sbox2{$#2\bullet$}%
    \ifdim\ht2<\ht0 %
      \ht0=\ht2 %
    \fi
    \sbox2{$\mathsurround=0pt #2#3$}%
    \hbox to \wd2{\hss\usebox{0}\hss}%
  }%
}

\def\XXint#1#2#3{{\setbox0=\hbox{$#1{#2#3}{\int}$ }
\vcenter{\hbox{$#2#3$ }}\kern-.58\wd0}}

\DeclareMathAlphabet{\mathpzc}{OT1}{pzc}{m}{it}
\DeclareMathAlphabet\mathpazo{OML}{zplm}{m}{it}
\DeclareMathAlphabet\matheuler{U}{eus}{m}{n} 
\DeclareMathAlphabet\mathat{T1}{ptm}{m}{n} 

\def \euA  {\matheuler{A}}

\def \euP  {\matheuler{P}}

\def \calA   {\mathcal{A}}

\def \calC   {\mathcal{C}}

\def \calF   {\mathcal{F}}

\def \calN   {\mathcal{N}}

\def \calX   {\mathcal{X}}

\def\mbf#1{\mathbf{#1}}	    
\def\mrm#1{\mathrm{#1}}	    
\def\mtt#1{\mathtt{#1}}	    

\def \mR     {\mathbb{R}}

\def \mP     {\mathbb{P}}

\def \mC     {\mathbb{C}}
\def \mP     {\mathbb{P}}   

\def \a      {\alpha}
\def \b      {\beta}
\def \g      {\gamma}
\def \d      {\delta}

\def \e      {\epsilon}
\def \s      {\sigma}

\def \th     {\theta}

\def \bxi    {\bm{\xi}}

\def \sb     {\mathbf{s}}

\DeclareMathOperator*{\Ran}{Ran}
\DeclareMathOperator*{\Ker}{Ker}
\DeclareMathOperator*{\Var}{Var}

\DeclareMathOperator*{\argmin}{arg\,min}

\DeclareMathOperator{\vect}{vec}

\def \supp   {\mrm{supp}}

\def \into   {\rightarrow}

\def \rsa    {\rightsquigarrow}

\def \cp     {\partial}  

\def \ssubset {\subset\!\subset}

\begin{document}

\begin{spacing}{1.1}

\title{\large{\textsc{sensitivity of regular estimators}}}

\author{\normalsize\textsc{yaroslav mukhin}\footnote{ ymukhin@mit.edu }}

\date{}

\maketitle
{\let\thefootnote\relax\footnote{{May 18, 2018}}}

\thispagestyle{articleStyle}

\begin{abstract}
This paper studies local asymptotic relationship between two scalar estimates.
We define sensitivity of a target estimate to a control estimate to be the directional
derivative of the target functional with respect to the gradient direction of the control
functional.
Sensitivity according to the information metric on the model manifold is the asymptotic covariance
of regular efficient estimators.
Sensitivity according to a general policy metric on the model manifold can be obtained from
influence functions of regular efficient estimators. 
Policy sensitivity has a local counterfactual interpretation, where the ceteris paribus
change to a counterfactual distribution is specified by the combination of a control parameter and
a Riemannian metric on the model manifold.
\end{abstract}

\pagestyle{articleStyle}

\section{Introduction}

Balancing simplicity of statistical methodology with complexity of economic modeling is a
challenge in empirical work. Structural models lead to estimators with nontransparent dependence
on data. Both structural and predictive models are subject to specification choices that have
nontransparent influence on inferences. However, regular estimators of parameters in these models
have simple asymptotic behavior and can be understood well locally. 
Regularity allows to draw local comparisons (approximations) between two estimators and
obtain local counterfactuals of their values.
For example, it may be useful to know that a structural estimator is locally well approximated
with a simple \{mean, variance, quantile, etc\}.
Or that two alternative specifications provide similar results not only at the sampling
distribution but in a neighborhood around it.
Sensitivity measures formalize local comparisons and counterfactuals, and add transparency
to inferences made with structural models.

We examine geometric foundations of estimator sensitivity and highlight the role
of the information metric in asymptotics of regular estimation.  Covariance of joint
asymptotic distribution is the information inner-product that measures alignment of
first-order approximations to regular parameters. This is a natural measure of local
approximation quality between two estimators.
Differentiability has a prominent role and a long history in regular asymptotics from von
Mises (1947) \cite{mises1947} to van der Vaart (1991) \cite{vaart1991differentiable} and
Newey (1994) \cite{newey1994asymptotic}, we go a step further and develop complete
differential calculus on the model.
We define sensitivity as a directional derivative and propose it as a general tool for local
counterfactual analysis as in Stock (1989) \cite{stock1989nonparametric} and 
Chernozhukov, Fern\'andez-Val and Melly (2013) \cite{chernozhukov2013inference}. 
Instead of specifying a counterfactual distribution of control variables, we think of
policy as shifting the value of a control parameter. Sensitivity measures the effect of
policy on the value of a target parameter. 
For example, the local effect of changing the \{mean, variance, quantile, etc\} of a
distribution on the \{mean, variance, quantile, etc\} of the distribution.
Both the implicit counterfactual distribution and the sensitivity (directional derivative)
depend on the way policy measures distances on the model.
Asymptotic covariance is shown to be such a directional derivative with a particular
choice of geometric primitives.

\vspace{0.1cm}

To put our work in perspective, let us disassemble empirical analysis in economics into a
stack of layers and interfaces.
At the top level, there is a model of economic quantities that are defined independently
of data.
This can be a structural model or a descriptive relationship between control and response
variables, say, quantity $\vartheta$ is of interest to the researcher.
For example, a price elasticity, a rate of return, a parameter of utility
function, a location or scale parameter.  At the bottom of the empirical analysis stack,
there are data from unknown distribution $P$ on sample space $(\calX,\calA)$ that can be
described with a statistical model $\euP$. 
For example, a random sample from a parametric, nonparametric or semiparametric model.
At the interface between the application and the data layers, high-level object $\vartheta$
is identified with a particular feature of the statistical model $\psi(P)$. 
Thus, the middle layer between data and application is a specification 
$\Psi:a\mapsto \psi_a$ that assigns a statistical parameter $\psi_a$ to the economic
quantity $\vartheta$ under modeling assumptions $a$ of the researcher, say, index set
$\euA$ describes all specifications entertained by the researcher.

Three logically independent types of variation in empirical inference about $\vartheta$ can be
distinguished based on the application, specification and data layer anatomy.
\emph{Application model sensitivity} analysis examines dependences within the mathematical
relationships of the application layer,  \cite[e.g.,][]{sobol1993sensitivity, 1404.2405}.
\emph{Specification sensitivity} arises from variations at the interface layer in mapping $\Psi$.
For example, $\vartheta$ can be
identified with a coefficient in a linear regression model or an \textsc{iv} equation, both
\textsc{ols} and \textsc{iv} can  be set up with different sets of covariates or instruments.  
Omitted variable bias is the quintessential example of variation in specification.
Both the statistical model $\euP$ and the unknown distribution $P$ of sampled data remain fixed
across different specifications, only the choice of statistical functional $\psi(P)$ that is used
for inference about $\vartheta$ changes.
Exploring specification variation for a fixed $P$ is analytically straightforward --
estimates of all interesting choices $\Set{\psi_a(P)}{a\in \euA}$ can be obtained, hopefully
uniform, inferences can be reported, a parametrization $\Psi$ can be differentiated with techniques
from calculus to find local effects of changing specification. 

\vspace{0.1cm}

This paper studies sensitivity of a fixed statistical functional defined on a statistical
model
$$
    \psi:\euP\into\mR
$$
to local variations of the data distribution $P$ within model $\euP$. We work strictly at the
data layer, holding specification fixed, but suggest both data level and application level
interpretations.  In mathematical terms, we consider differential calculus of functionals on the
model manifold under different Riemannian geometries.
\emph{Statistical model sensitivity} is a directional derivative of the statistical functional. 
Since a typical statistical model behind economic
applications is an infinite-dimensional space, it is helpful to identify a direction on the model
with a tractable statistical parameter, denoted $\nu(P)$.
Sensitivity with respect to parameter $\nu(P)$ is the partial derivative along its
gradient vector $\nabla\nu$, denoted by $\cp_\nu$ operator:
$$
\cp_\nu \psi \coloneqq \lim_{h\into 0} h^{-1} 
\big[
    \psi(P + h \cdot \nabla\nu) - \psi(P) 
\big].
$$

Practical utility of sensitivity analysis comes from the fact that it is closely related to
asymptotic approximations for a large class of estimators.
The main observation is that influence functions are gradients according to the information
geometry of the model. Gradients in any other geometry on the model are linear
transformations of influence functions. By varying geometric primitives in the definition
of sensitivity $\cp_\nu\psi$, researcher obtains different local counterfactual values of
$\psi$, corresponding to different perturbations on $\euP$ that change $\nu$ in a
controlled way. One of such counterfactual is given by the asymptotic covariance of two
regular estimators.

For a pair of estimators on statistical model $\euP$ with standard asymptotic behavior
%
\begin{align} \label{eqn:asy}
    \sqrt{n} \begin{pmatrix} 
			\wh{\psi}_n - \psi \\
			\wh{\nu}_n - \nu 
	     \end{pmatrix}   \rightsquigarrow (\wt{\psi},\wt{\nu}) \sim
    N(0, \Sigma ),
      \text{ where } \Sigma = 
	\begin{bmatrix} 
	    \sigma_{\psi\psi} & \sigma_{\psi\nu} \\
	    \sigma_{\psi\nu} & \sigma_{\nu\nu}
	\end{bmatrix},
\end{align}
%
%
\vspace{-14pt}
\begin{defn} \label{def:estSens}
    the \emph{estimator sensitivity} of $\psi(P)$ to $\nu(P)$ is
    \begin{align*}
	\Lambda & \coloneqq \sigma_{\psi\nu} / \sigma_{\nu\nu}
	    \qquad\quad\;\;
	    \text{the coefficient of }
	    E[\, \wt{\psi} \,\vert\, \wt{\nu} \,], 
	\\
	\shortintertext{\noindent and \emph{estimator sufficiency} of $\nu(P)$ for $\psi(P)$ is }
	\Delta & \coloneqq  
	    \sigma^{2}_{\psi\nu} / \sigma_{\nu\nu} \sigma_{\psi\psi}
	    \qquad
	    \text{the } 
	    R^2 = \Var \big( E[\,\wt{\psi}\vert\wt{\nu}\,] \big) / \Var \big( \wt{\psi} \big).
    \end{align*}
\end{defn}
\vspace{-16pt}

\noindent
The $\Lambda,\Delta$ measures were introduced by Gentzkow and Shapiro (2015) \cite{Gentzkow2015}
for the purpose of comparing a nontransparent estimator $\wh\psi_n$ to a tractable statistic
$\wh\nu_n$.
Andrews, Gentzkow and Shapiro (2017) \cite{andrews2017measuring} interpreted $\Lambda$ as a
measure of local specification sensitivity of \textsc{gmm} functionals implicitely parametrized by
the population value of moments $E g(X, \psi_a(P) ) = a$.\footnote{From the fact that Jacobian $G$
of moments does not depend on specification parameter $a$, it follows that dependence of moments
on $a$ must be additive.}

We define sensitivity directly on the model using techniques of differential geometry, rather than
in terms of the asymptotic distribution of estimators as in \cite{Gentzkow2015}.
We then relate our sensitivity of functionals to asymptotic distributions of estimators using
results from semiparametric efficiency theory. 
This relationship is similar to Newey (1994) \cite{newey1994asymptotic},
but in our definition we allow for an explicit choice of geometric primitives.
We show that, in information geometry of $\euP$:
Estimator sensitivity $\Lambda(\wh\psi,\wh\nu)$ is (i) the directional derivative $\cp_\nu\psi$ of
$\psi(P)$ in the direction of $\nu(P)$.
Estimator sufficiency $\Delta(\wh\psi,\wh\nu)$ is (ii) the square of cosine of the angle made by linear
approximations to $\psi$ and $\nu$ at $P$, (iii) the relative size of partial derivative of $\psi$
along $\nu$ to total derivative of $\psi$, (iv) the efficiency gain in estimating $\psi$ obtained
by fixing population value of $\nu$. With other geometries on $\euP$, measures (i-iii) are
available but not reflected in the asymptotic distribution of estimators.

Our investigation is inspired by \cite{Gentzkow2015} but we proceed in a different direction from
their line of inquiry.
The main objective of this paper is to provide interpretation of $\Lambda,\Delta$ measures from
semiparametric efficiency perspective. This leads us to information geometry and motivates our
local counterfactual interpretation of sensitivity, which we generalize by allowing a policy
metric instead of the intrinsic information metric of the geometry behind statistical
efficiency. 
Apart from generalizations, our inquiry fundamentally diverges from \cite{Gentzkow2015} in that
we make a clear distinction between varying specification $a\mapsto\psi_a$, holding $P$ fixed, and
varying distribution $P$, holding specification $a\mapsto\psi_a$ fixed, and consider only the
latter exercise. By contrast, \cite{andrews2017measuring,Gentzkow2015} are primarily concerned
with variation in the specification of moment conditions in \textsc{gmm} functionals,
which are \emph{not} deviations on the statistical model.
This paper and \cite{andrews2017measuring,Gentzkow2015}  obtain complementary interpretations
for quantities $\Lambda,\Delta$ which should only increase their value in practice.

\vspace{0.1cm}

We suggest two types of applications of statistical model sensitivity. A data level
interpretation as a measure of local alignment of two functionals can be used to
compare competing specifications or target specifications to tractable statistics.  An
application level interpretation as a derivative can be used for local 
counterfactual analysis and policy evaluation.

Measures (ii-iv) above quantify the quality of local approximation of $\psi$ by $\nu$ in a
neighborhood of $P$.
Linear approximation of $\psi$ determines first order asymptotic
behavior of estimates $\wh\psi$.  Sensitivity thus provides an analytic tool for exploring
inferences based on the asymptotic distribution of estimates of $\psi$. Reporting sensitivity to
tractable parameters $\nu$ helps explain how inferences about $\psi(P)$ are obtained from
$P$. 
See \cite{andrews2017measuring,Gentzkow2015} and references therein for a discussion on
transparency and empirical examples.
In the case with multiple specifications for $\vartheta$, the natural course is to report all
estimates $\Set{\wh\psi_a}{a\in \euA}$. This provides a one-point comparison of different
specifications at the sampling distribution $P$.  
Reporting $\psi_a(P)$ similar to $\psi_b(P)$, positive estimator sensitivity
$\Lambda(\wh\psi_a,\wh\psi_b)$ and estimator sufficiency $\Delta(\wh\psi_a,\wh\psi_b)$ close to
one, can be offered as formal evidence that results are not sensitive to specification in a
\emph{neighborhood} of $P$.  We call these applications estimator or information sensitivity.
\footnote{
Note that identification and consistency of estimates $\wh\psi$ are global properties of the
functional and the model and thus are outside of the scope of local sensitivity analysis.
}

Directional derivatives (i) provide a simple description of the local behavior of functional $\psi$ at
distribution $P\in\euP$. 
For streams of random samples generated by $P_h = P + h \, \wt\nu_P$,
where $\wt\nu_P$ is the gradient of $\nu(P)$, the limits under $P_h$ of estimators $\wh\psi$ and
$\wh\nu$ are:
$$
\wh\psi 
\xrightarrow[]{P_h}
\psi(P) + h \cdot \cp_\nu\psi + o(h)
\quad\text{and}\quad
\wh\nu 
\xrightarrow[]{P_h}
\nu(P) + h \cdot \cp_\nu\nu + o(h)
.
$$
We see that sensitivity $S(\psi,\nu)\coloneqq \cp_\nu\psi /\cp_{\nu}\nu $ is the local effect
on the value of $\psi(P)$ of a ceteris paribus change in the value of $\nu(P)$ accomplished by
changing the underlying distribution from $P$ along $P_h$.  
This is the local version of the
counterfactual analysis that typically takes $\psi(P)$ to be some location parameter of a
response variable $Y$ and $\nu(P)$ to be the marginal distribution of a policy variable $X$
\cite[e.g.][]{stock1989nonparametric,heckman2007econometric,chernozhukov2013inference}.
Finally, one can use the identification of statistical functionals $\psi,\nu$ with economic
quantities $\vartheta,\eta$ of the application layer and interpret the local relationship
$S(\psi,\nu)$ as the partial derivative of $\vartheta$ with respect to $\eta$.
We call these applications policy sensitivity and argue that it should be based on a geometry of
$\euP$ with a policy metric motivated by the application, rather then the information metric
dictated by technicalities of asymptotic approximations.

Policy metric is a local notion distance on the model $\euP$. Asymptotic inference implicitly
relies on the information metric that measures ``statistical'' distances on the model. Metric
determines the direction $\wt\nu$ on the model along which policy shifts in the value of $\nu$ are
achieved. Thus, the combination of control functional $\nu$ and policy metric determines the path
of counterfactual distributions $P_h$ along which sensitivity of target functional $\psi$ is
measured. We describe a simple procedure for specifying and interpreting policy metrics, and
illustrate the analysis with a Monte Carlo experiment.
Parametrizing directions on the model by a control functional and a policy metric is a tractable
and flexible way to reason about local counterfactuals.

\vspace{0.1cm}

The scope and contribution of this paper is to provide geometric foundation for
statistical model sensitivity analysis and to highlight the importance of the metric of the model. 
We provide new geometrically motivated methodology for counterfactual analysis. This appears to be
a novel use of geometry in econometrics and statistics. More specifically,
we introduce the notion of a policy metric on a statistical manifold, including semiparametric and
nonparametric models.  We then define sensitivity as a directional derivative with respect to
policy gradient of a control statistical parameter.
This geometric formulation enables us to interpret policy sensitivity, including the covariance of
asymptotic distribution, as a local counterfactual. 
In order to compute and estimate policy sensitivities, we obtain a result that relates policy
gradients to influence functions. We provide high level conditions for consistency of estimated
sensitivity.  Detailed econometric analysis of estimation and inference for real-valued and
distributional local counterfactuals is left to future work.

This paper draws on and contributes to several seemingly unrelated literatures. 
Geometric foundations in statistical inference have been investigated by many authors:
Hotelling (1930) \cite{hotelling1930} considers the spaces of statistical parameters as curved
surfaces embedded in Euclidean space, one of which can be seen in \Cref{fig:hyperbolicNormal}.
Mahalanobis (1936) \cite{mahalanobis1936generalized} defines general distances between statistical
populations and notes parallels with special relativity.
Rao (1949) \cite{rao1949appendix} writes down the information metric of a population space
(parametric model) in local coordinates and describes geodesics between two distributions.
Amari (1985, 2000) \cite{amari1985,amari2000} provides geometric insight into asymptotic
efficiency in parametric models. To this literature we contribute by applying differential
geometry to infinite-dimensional models and by new methodology motivated by geometry.
Specification sensitivity analysis based on $\Lambda,\Delta$ was introduced by Gentzkow and
Shapiro (2015) and Andrews, Gentzkow and Shapiro (2017) \cite{Gentzkow2015,andrews2017measuring}.
Semiparametric efficiency theory shows that variance of asymptotic Gaussian distribution  in large
statistical models is the information norm of the differential e.g.
Stein (1956) \cite{stein1956}, 
Koshevnik and Levit (1976) \cite{Koshevnik76},
Pfanzagl (1982) \cite{pfanzagl1982contributions}, 
van der Vaart (1991) \cite{vaart1991differentiable},
Bickel et al. (1993) \cite{bickel1993efficient}
but does not make explicit use of modern geometry.
We contribute to the efficiency literature by modelling large models as manifolds.

We organize the paper as follows:
In \Cref{sec:metricSensitivity} we define sensitivity using econometrics language of
semiparametric efficiency and provide a Monte Carlo example to illustrate the methodology.
To make geometric ideas of this paper accessible without requiring familiarity with Riemannian
geometry and semiparametric efficiency, we consider in \Cref{sec:surfaces} the special case of a
two-dimensional statistical model embedded in $\mR^3$. This allows a graphical illustration of
methodology and explicit calculations.
In \Cref{sec:geom} we review required foundations from differential geometry, state the general
definition of sensitivity measures, explain how they depends on geometric primitives of the model, 
and discuss analytic interpretation of these measures.
In \Cref{sec:sensRegEst} we apply results of semiparametric efficiency theory to obtain
information sensitivity from regular efficient estimators, relate policy gradients to influence
functions, and briefly consider consistency of estimated policy sensitivity.
We work out some simple examples in \Cref{sec:examples} and give a self-contained summary of
efficiency theory results we cite in \Cref{sec:appendix}.

\section{Econometric Introduction to Sensitivity} \label{sec:metricSensitivity}

This section provides an informal introduction to sensitivity, explains how it relates to
geometry of the statistical model and shows how to compute sensitivity for tractable policy
metrics.
We provide an axiomatic development and technical details in \Cref{sec:geom,sec:sensRegEst}, and
focus on the main ideas below, all calculations are deferred to \Cref{sec:examples}.

Let $\euP$ be a statistical model.
We are interested in estimating
parameter $\psi:\euP\into\mR$ or, possibly, a set of alternative specifications
$\Set{\psi_a:\euP\into\mR}{a\in \euA}$ defined on the same model.
Statistical functionals estimable at the parametric rate $\sqrt{n}$ are smooth. Therefore we can
define sensitivity as a directional derivative of $\psi$ along a tangent vector $v$ to the model
$\euP$ at the sampling (true) distribution $P_0$. Tangent vector $v$ is the score of a
one-dimensional parametric submodel $t\mapsto P_t$ defined in a neighborhood of $0\in [0,\e)$:
\begin{align*}
    v(x) = \frac{d}{dt}_{|t=0} \log dP_t(x).
\end{align*}
For the purposes of interpreting sensitivity, score $v$ stands for any submodel that
satisfies above derivative condition in quadratic mean. All such submodels admit the same local
counterfactual interpretation of sensitivity.  The collection of different scores $v$, obtained
from all smooth submodels through $P_0$, is called the tangent set, denoted $T_{P_0}\euP$. On a
fully nonparametric model, the tangent set is the space $L^2_0(P_0)$ of $P_0$ square-integrable
functions with zero mean.  Parametric and semiparametric models restrict the tangent set in
significant ways. Because we are not concerned with efficiency here, we can assume that the
tangent set is unrestricted.

Sensitivity of $\psi$ along the tangent vector $v\in L^2_0(P_0)$ is the local effect
of changing the distribution in the direction of score $v$:
\begin{align*}
    \cp_v\psi \coloneqq \lim_{t\into 0} t^{-1}
    \big[ 
	\psi(P_0 + tv) - \psi(P_0)
    \big] 
    .
\end{align*}
Here the perturbation $P_0+tv$ is understood to be any one-dimensional submodel $P_t$ with
score $v$.  For example, $dP_t = (1 + tv)dP_0$ or $dP_t=c(t)\exp(tv)dP_0$. To compute sensitivity
we can use the influence function of $\psi$:
\begin{align*}
    \cp_v\psi = \int_\calX \wt\psi v \, dP_0
    .
\end{align*}

As defined above, sensitivity is not very useful. The problem is that tangent space $T_{P}\euP$ 
typically does not have an obvious parametrization that would enumerate
all scores and put different sensitivities into context of the application layer.
To make sensitivity analysis convenient for the practitioner, the direction $v$ should be
associated with a tractable parameter of interest to the researcher. 
This can be a statistical functional motivated by the
application layer, e.g. a related economic quantity or an alternative specification of the same
quantity. Or this can be a data level parameter that provides a tractable summary of distribution
$P_0$, e.g. a mean or a quantile. We call this parameter a control functional and denote it by
$\nu:\euP\into\mR$.

The natural direction to associate with $\nu(P)$ is the gradient where functional increases most
rapidly. This is analogous to the way Cartesian coordinates work, if we think of
coordinates as functions of the point.
However, it is not enough to pick a control functional to specify the direction of sensitivity.
This should not be surprising, because $T_{P}\euP$ is a large space, for which we have not
introduced any structure.

Gradients depend on the notion of distance on the model $\euP$. A metric at $P\in\euP$ is an
inner-product norm $\norm{\cdot}_P$ on tangent vectors $T_P\euP$. The distance between $P_0$ and $P_\e$
along submodel $P_t$ is the sum of lengths of tangent vectors along the curve:
\begin{align*}
    \mrm{dist}^2(P_0,P_\e) = \int_0^\e \norm{\tfrac{d}{dt}_{|t=h}\log dP_t}^2 dh
    .
\end{align*}
Different metrics define different distances on $\euP$ and generate the different geometries.

Influence function $\wt\nu$ is the gradient of $\nu$ according to the information geometry of
$\euP$ that has metric $\norm{v}^2_P = \int v^2 \, dP$. Information $\norm{v}_{L^2(P)}$
measures statistical discrepancy between $P$ and a perturbation $P+\e v$ in the direction of score
$v$. 
Influence function is the direction on the model along which change in the value of
the functional is greatest per statistical deviation away from $P$. This direction is least
favorable on the model for estimating $\nu$ from random samples of $P$.

Calculation of influence functions is a standard exercise in efficiency literature, we refer to
Ichimura and Newey (2015) \cite{ichimura2015influence} for a modern treatment and use their
formula as a convenient definition:
\begin{align} \label{eqn:influenceFct}
    \wt\psi(z) \coloneqq \lim_j \left[ \tfrac{d}{dt} \psi(P_{z,t}^j)_{\big|t=0} \right].
\end{align}
Let us fix a simple example. Let the target functional be a generic moment of data 
$\psi_\rho(P) = \int\rho(x)dP$, and let the control functional be a quantile of data 
$\nu_\tau(P) = F^{-1}_{X(j)}(\tau)$.  The mean and the $\tau$-quantile have influence functions
$$
\wt\psi_\rho(x) = \rho(x) - \psi_\rho(P)
\quad
\text{and}
\quad
\wt \nu_\tau (x)=\dfrac{\tau - 1_{[x_i,\infty)}(\nu_\tau (P))}{f_{X(i)}(\nu_\tau(P))}
.
$$
The information sensitivity of the mean to the quantile
\begin{align*}
    \frac{\cp\psi_\rho}{\cp \nu_\tau} \coloneqq
    \frac{1}{f_{X(i)}(\nu_\tau)} 
    \int 
    \big[
	\rho(x) - \psi_\rho
    \big]
    \big[
	\tau - 1_{[x_i,\infty)}(\nu_\tau)
    \big] 
    \, dP_0(x)
\end{align*}
is the asymptotic covariance of regular efficient estimators $\wh\psi,\wh\nu$.

To interpret this, rescale $\Lambda(\psi,\nu)\coloneqq \cp_\nu\psi / \norm{\wt\nu}^2_P$ and recall
the original definition of information (in infinite-dimensional models) form Koshevnik and Levit
(1976) \cite{Koshevnik76}: $\Lambda$ is the effect on the mean $\psi(P_0)$ of a perturbation to
$P_0$ along a one-dimensional submodel $P_h$ that satisfies two requirements:
\begin{enumerate}[label=($\Lambda$-\roman*), itemsep=0cm, topsep=0.0cm, itemindent=0.5cm, align=left]
    \item
	generate an increment $h$ in the value of quantile $\nu_\tau$, so that 
	$\nu_\tau(P_h)=\nu_\tau(P_0) + h$ ;
    \item
	minimize the information distance between $P_0$ and $P_h$.
\end{enumerate}
Information sensitivity $\Lambda$ measures the effect of this perturbation on the counterfactual
value of the mean:
$$
\psi_\rho(P_h)= \psi_\rho(P_0) + h \Lambda(\psi_\rho,\nu_\tau) + o(h).
$$

Sensitivity to perturbations along the least favorable submodel is interesting for comparing
statistical properties of estimators. For example, if $\psi,\nu$ are two alternative
specifications for the same economic quantity, then information sufficiency 
$\Delta(\psi,\nu)\coloneqq \abs{\cp_\nu\psi}^2/\norm{\wt\psi}^2_P\norm{\wt\nu}^2_P$ is a natural
measure of local similarity of the two estimates.
But the choice of least favorable submodel as the counterfactual distribution when measuring
the response in $\psi$ to changes in $\nu$ has no structural or causal foundation. Our point
is to make this choice explicit.

A general sensitivity of parameter $\psi$ can thus be specified by a combination of:
\begin{enumerate}[label=(S-\roman*), itemsep=0cm, topsep=0.0cm, itemindent=0.5cm, align=left]
    \item \label{item:policySens1}
	control functional $\nu$ whose value is being manipulated;
    \item
	metric $\norm{\cdot}_P$ on the tangent space $T_{P}\euP$ that determines the direction of the
	one-dimensional submodel along which control functional changes most rapidly.
\end{enumerate}
To contrast general sensitivity with information sensitivity, we will call the metric used to
determine gradients a \emph{policy} metric, the direction along which the sensitivity is measured
a \emph{policy} gradient, and the directional derivative itself a \emph{policy} sensitivity.
Control functionals and a policy metric provide a partial parametrization of the tangent space
$T_P\euP$ that enables local counterfactual analysis motivated by the application.

A tractable way to specify a policy metric is to postulate a policy distribution $Q_P$ whose density
function $dQ_P(x)$ reflects the cost of displacing a unit of mass at location $x$ in the sample
space.  The choice $Q_P$ should be motivated by the application.
The resulting policy metric is $\norm{v}_{L^2(Q_P)}^2 = \int \abs{v}^2dQ_P$.
Policy sensitivity with this metric is
\begin{align*}
    S_\nu\psi \coloneqq
    \int_\calX \wt\psi \,  \nabla\nu \, dP_0 \; / \; \norm{\nabla\nu}^2_{L^2(Q)}
    ,
\end{align*}
where the scaled gradient $v = \nabla\nu / \norm{\nabla\nu}^2_{L^2(Q)}$ is the score 
$\tfrac{d}{dh}_{|h=0} \log dP_h$ of a one-dimensional submodel $P_h$ that solves the following
program for a sufficiently small $\e$:
\begin{align*}
    \min_{(0,\e) \ni h \mapsto P_h}
    \int_0^\e dh
    \int 
    \left[
	1 - \frac{dP_h^{1/2}}{dP_0^{1/2}}
    \right]^2 \; dQ + o(\e)
    \qquad \text{s.t.}  \qquad
    \nu(P_h) = \nu(P_0) + h + o(h).
\end{align*}
Under some regularity conditions, policy gradient of functional $\nu:\euP\into\mR$ with respect to
policy metric $\norm{\cdot}_{L^2(Q)}$ is 
\begin{align*}
    \nabla \nu 
    =
    \Big[\wt\nu -  P\wt\nu\tfrac{dP}{dQ} / P \tfrac{dP}{dQ}  \Big]\tfrac{dP}{dQ}.
\end{align*}
The effect of changing the metric from information to policy is very intuitive: the influence
function is rescaled by the likelihood ratio of information to policy and recentered.
Policy sensitivity measures the effect on the counterfactual value of target functional $\psi$
from the perturbation to the value of control functional $\nu$ along any submodel with policy
gradient $\nabla\nu$:
$$
\psi(P_h)= \psi(P_0) + h S(\psi,\nu) + o(h).
$$

\subsection{Monte Carlo example.}
%
\begin{figure}[h]
\centering
\begin{minipage}{0.49\textwidth}
\includegraphics[scale=0.55, trim=0in 0in 0in 0in, clip=flase ]{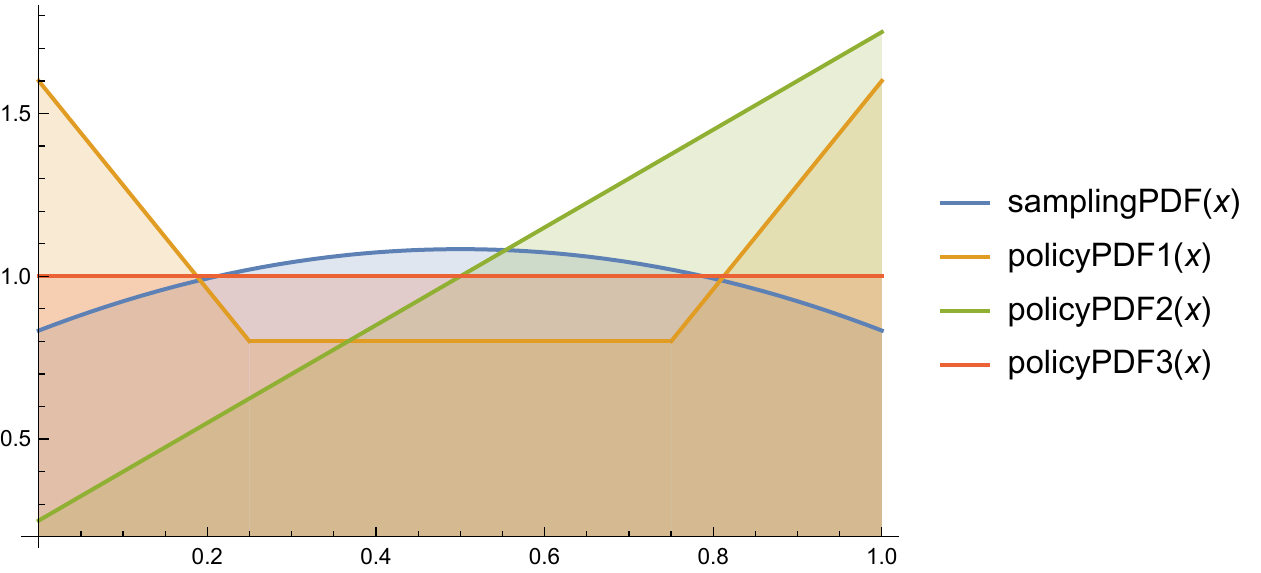}
    \subcaption{Policy distributions}
    \label{fig:policyPDFs}
\end{minipage}
\begin{minipage}{0.49\textwidth}
\includegraphics[scale=0.55, trim=0in 0in 0in 0in, clip=false ]{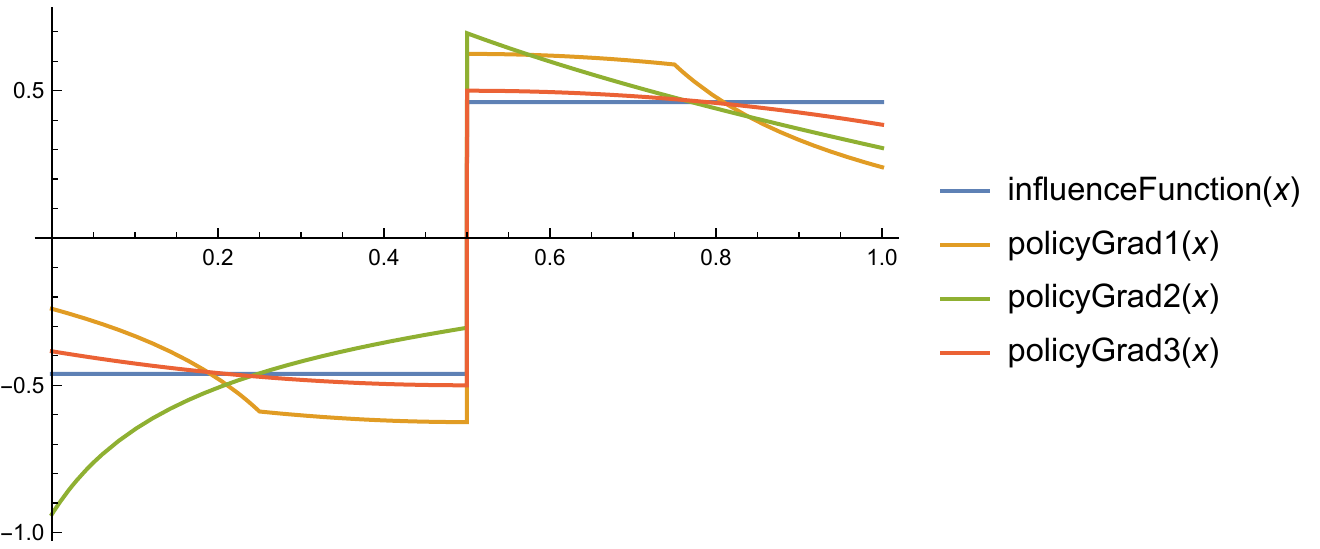}
    \subcaption{Policy gradients of the median $\nu$}
    \label{fig:policyGradients}
\end{minipage}
    \caption{}
    \label{fig:policyPerturbations}
\end{figure}

Let $X,Y$ be continuously distributed according to joint distribution $P$ on the interval
$[0,1]^2$, and suppose that $Y$ is a measure of income, $X$ is a measure of education.
Application layer postulates that $Y$ is a response variable, whereas $X$ is a control variable of
intereset.
Let the target and control functionals
$$
\psi(P) = \int_{[0,1]^2} y dP
\qquad \text{and} \qquad
\nu(P) = F_X^{-1} \big(\tfrac{1}{2} \big)
$$
be the mean of response variable $Y$ and the median of control variable $X$.
In our simulation, we take 
$$
Y|X \sim  \mrm{Beta}(\a,\b)
\quad \text{with} \quad
\a=2, \;\b=5-5X
\quad \text{so that} \quad
E[Y|X] = 2/(7-5X)
$$
the conditional mean of income given education is positively correlated with education. Marginal
distribution of $X$ is shown on \cref{fig:policyPDFs}, marked
{\footnotesize\textsf{samplingPDF(x)}}.

We are interested in the predictive effect on income, via target functional $\psi(P)$, of a policy
that perturbs the marginal distribution of education. It is assumed that the perturbation does not
change the conditional distribution $P_{Y|X}$.
Policy is designed to increase the median level of education $\nu(P)$  by some prescribed amount
($0.1$ in the simulation).  Three implementations of policy are proposed. 

\begin{figure}[h]
\centering
\begin{minipage}{0.6\textwidth}
\includegraphics[scale=0.7, trim= 0in 0in 0in 0in, clip=false, ]{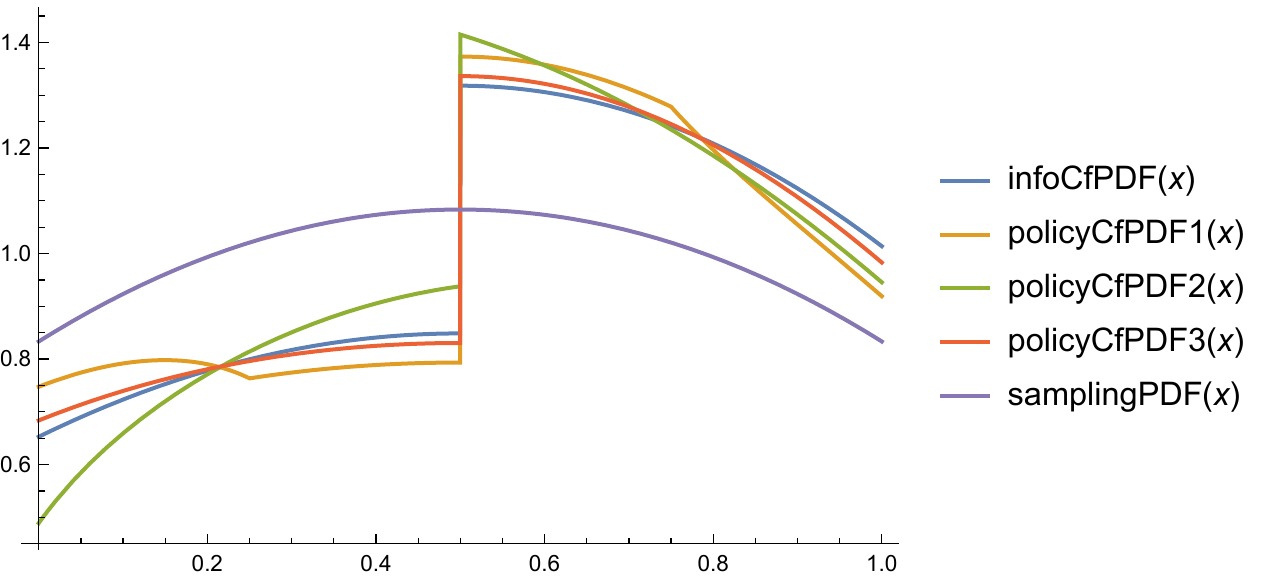}
\subcaption{Counterfactual distributions}
\label{fig:policyCounterfactualPDFs}
\end{minipage}
\begin{minipage}{0.39\textwidth}
    \vspace{0.5cm} \hspace{1cm}
\begin{tabular}{ c c }
    Metric & Sensitivity \\
    \hline\noalign{\smallskip}
   $L^2(P_X)$ & 0.304102 \\
   $L^2(Q_1)$ & 0.251316 \\
   $L^2(Q_2)$ & 0.283477 \\
   $L^2(Q_3)$ & 0.289154 \\
\end{tabular}
    \vspace{0.5cm}
\subcaption{Sensitivity of mean $\psi$ to median $\nu$}
\label{table:policySensitivities}
\end{minipage}
    \caption{Local counterfactuals}
    \label{fig:policyCounterfactuals}
\end{figure}

$P_X:$
The perturbation along the least favorable submodel in the direction of the influence function of
the median $\nu$ has the effect $\Lambda(\psi,\nu) = 0.3041$ on the mean $\psi$. The influence
function is marked {\footnotesize\textsf{influenceFunction(x)}} in \cref{fig:policyGradients}.
The density function of the counterfactual distribution $dP_h = (1+h \wt\nu)dP$
that produces $\nu(P_h)\approx 0.6$ is marked {\footnotesize\textsf{infoCfPDF(x)}} in
\cref{fig:policyCounterfactuals}. The information counterfactual value of the mean is 
$\psi(P_h)\approx \psi(P) + 0.3041 \times 0.1$.

$Q_1:$ 
The first policy proposal minimizes the taxpayers' cost of policy.
It is argued that increasing the proportion of highly educated workers and reducing the
proportion of workers with most basic education is progressively more costly as one approaches the
extremes of the distribution. This may be due to higher investment requirements of
displacing workers at the extremes. This proposal is summarized with policy cost density function
$dQ_1$, marked {\footnotesize\textsf{policyPDF1(x)}} in \cref{fig:policyPDFs}. Distribution $Q_1$ 
defines policy metric $\norm{\cdot}_{L^2(Q_1)}$ on deviations from sampling distribution of
education $P_X$ and produces a policy gradient function $\nabla_{Q_1}\nu$, marked
{\footnotesize\textsf{policyGrad1(x)}} in \cref{fig:policyGradients}. The resulting counterfactual
distribution, marked {\footnotesize\textsf{policyCfPDF1(x)}} in
\cref{fig:policyCounterfactualPDFs}, is closer to the original sampling distribution $P_X$ below
the first and above the third quartiles, and further away at the interquartile range, compared to
the information counterfactual. The counterfactual value of the median is
$\nu(P+h\nabla_{Q_1}\nu) \approx 0.6$, and the policy sensitivity is $S_{Q_1}(\psi,\nu) = 0.2513$,
so the counterfactual value of the mean is 
$\psi(P+h\nabla_{Q_1}\nu)\approx \psi(P) + 0.2513\times 0.1$.

$Q_2:$
The second policy proposal minimizes economic inequality by designing the perturbation to have the
strongest effect at the  lowest levels of education and tapering off toward the highest levels of
education. This is achieved with policy distribution $dQ_2$, marked
{\footnotesize\textsf{policyPDF2(x)}} in \cref{fig:policyPDFs}, and confirmed by the
counterfactual distribution marked  {\footnotesize\textsf{policyCfPDF2(x)}} in
\cref{fig:policyCounterfactualPDFs}.  The sensitivity $S_{Q_2}(\psi,\nu) = 0.2835$ fits in between
the information sensitivity $\Lambda$ and the $Q_1$ policy sensitivity $S_{Q_1}$. This is
explained by noting that the mean of $Y$ is positively related to the mean of $X$, and that
deviations with more mass at the tails effect the mean stronger than deviations that displace more
mass around the median of the distribution.  

$Q_3:$
The third policy proposal minimizes the macroeconomic shock by assigning equal cost to deviations
across all levels of education.
Perturbation profile, the gradient $\nabla_{Q_3}\nu$, under policy metric $\norm{\cdot}_{L^2(Q_3)}$
is most similar to the influence function $\wt\nu$ of the information metric
$\norm{\cdot}_{L^2(P_X)}$. 
This is because both the sampling distribution and the policy measure $Q_3$ are relatively flat.
The similarity is reflected in the counterfactual distributions and sensitivities as well.

Counterfactual value of $\nu$ in each case is approximately $0.6$.
We compute the sensitivity of $\psi$ to changes in $\nu$ under each of the four counterfactual
distributions and report results in \cref{table:policySensitivities}.

\begin{remark}
    The control functional determines the overall profile of the perturbation to $P$. 
    The distribution in the policy metric determines the intensity with which the perturbation is
    applied across the sample space with higher policy density attenuating the perturbation and lower
    policy density intensifying the perturbation.
\end{remark}

\subsection{Sensitivity of GMM.}

In this section we illustrate sensitivity analysis with \textsc{gmm} and descriptive statistics.
We consider \textsc{gmm} functionals on the nonparametric model $\euP$ that is constrained only by
regularity (smoothness, integrability) conditions.  Application layer provides a parameter space
$\Theta\subset\mR^p$ for the economic quantity of interest $\vartheta$ and a vector of moment
criterion functions
$
g:\calX\times\Theta \into \mR^r,
$
assumed to be sufficiently smooth in parameter $\th$ and sufficiently integrable over the
sample space $\calX \subset \mR^d$. Integrals with respect to distribution $P$ of data are written
as $Pg(\th) = \int_\calX g(x,\th)\,dP(x)$.
It is assumed that the economic quantity $\vartheta$ is ``over-identified'', meaning that $r>p$,
and that $G\coloneqq P\cp_\th g(\th)$ and $\Omega\coloneqq P g(\th)g(\th)^T$ are full rank at
$\th=\psi(P)$.

Researcher \emph{specifies} that the value of $\vartheta\in\Theta$ is given by a function
$\psi:\euP\into\Theta$ of the statistical model. \textsc{gmm} estimation is set up from the
application layer assumptions that 
\begin{align}\label{eqn:moment_assumptions3}
    P g(\vartheta) = 0 \tag{$a_M$}.
\end{align}
These assumptions are typically optimality conditions of the interactions described by the
application layer model or postulated by the researcher orthogonality conditions.
Often these models are highly stylized and are not expected to
describe real-world data precisely.
Our view is that \cref{eqn:moment_assumptions3} assumptions should not be taken literally
to data, that the role of specification is nontrivial and deserves attention (but not our
focus here).
\textsc{gmm} functionals are defined by
$$
\psi_W(P) \coloneqq \argmin_{\theta\in\Theta} P g(\theta)^T W Pg(\theta).
$$
In the over-identified case, weighting determines the functional and should be chosen based on
application layer considerations. We consider only deterministic positive definite weighting
matrices for now. 
We compute a set of information sensitivities to compare a given \textsc{gmm} functional $\psi_W$
to tractable summaries of the data and to  alternative specifications $\psi_{A}$ obtained by using
a different weighting.
As directions we use descriptive statistics such as quantiles 
$q_\tau  \coloneqq F_{X(i)}^{-1}(\tau)$  and generic moments $\nu_\rho(P)\coloneqq
P\rho(X)$ of the data. Here the moment function $\rho:\calX \into \mR$ can be, for example,  a
component $\rho(x)=x_i$ of the data or a component of the moment criterion vector $\rho(x) =
g_i(x,\psi_W)$ .

Information sensitivity is simple to compute and offers greater insight into inferences based
on asymptotic approximations. Consider the \textsc{gmm} functional. The economic model that leads
to formulation of functional $\psi_W$ may be complicated, but the asymptotic distribution
of estimates, and inferences derived from it, are completely determined by the local behavior of the
functional at $P$. Information sensitivities and the complementary sufficiency measures provide tractable
one-dimensional summaries of this local variation:
$$
\cp_\nu \psi_W = P \wt\psi_W \wt\nu
\quad
\text{and}
\quad
R(\psi_W,  \nu) =   (P \wt\psi_W \wt\nu)^2 /  P \wt\psi_W^2 P \wt\nu^2
.
$$
Information sufficiency is an $R^2$ statistic that indicates how well the control functional
$\nu$ approximates local variation of the target functional $\psi_W$. Specifically, $R$ is the
square of cosine of the angle made by tangent hyperplanes to $\psi$ and $\nu$.
If $R(\psi_W,\nu)$ is close
to one, then inferences based on asymptotic approximations around $\psi_W(P)$ are obtained from
$P$ in the same way as inferences about $\nu(P)$. By making local comparisons of complicated
structural functionals $\psi_W$ to simple features of the data $q_\tau, \nu_\rho$, the statistical
part of the empirical analysis can be made transparent \cite{Gentzkow2015,andrews2017measuring}.
Another application is to compare two competing specifications $\psi_W$ and $\psi_A$ locally in
the neighborhood of $P$. Reporting $R(\psi_W,\psi_A)$ close to one can be offered as formal
evidence that the choice of weighting does not change results in a neighborhood of $P$.
Conversely, observing $R(\psi_W,\psi_A)$ close to zero warrants careful examination of
specification.


Asymptotic distribution of \textsc{gmm} estimators on misspecified models has been investigated by
Imbens (1997) \cite{imbens1997one}, Hall and Inoue (2003) \cite{hall2003large}, we derive the
influence function and policy gradients of the functional in order to provide sensitivity
analysis.
The influence function of the \textsc{gmm} functional on  a fully nonparametric model where moment
conditions \eqref{eqn:moment_assumptions3} are possibly violated is
\begin{align*}
    \wt\psi_W
    =
    - \Big[
	(Pg(\th)^T W \otimes I_p) \;\;
	\cp_\th \vect \big[ ( \cp_\th g(\th) )^T \big]
	+
	P[\cp_\th g(\th)]^T W P[\cp_\th g(\th)]
    \Big]^{-1} \times
    \qquad
    \\
    \quad \times
    \Big( 
	(Pg(\th)^T W \otimes I_p) \;\;
	\vect \big[ ( \cp_\th g(\th) )^T \big]
	+
	P[\cp_\th g(\th)^T] W \; g(\th) 
    \Big).
\end{align*}
The sign of sensitivity $\cp\psi_{W,i} /\cp\psi_{A,i} = P \wt\psi_{W,i} \wt\psi_{A,i}$ shows if
the two specifications for $\vartheta_i$ move in the same direction at $P$, and sufficiency 
$R(\psi_{W,i},\psi_{A,i})$ quantifies the alignment of two specifications locally at $P$.
Furthermore, sufficiency $R(\psi_{W,i}, \nu_{g(j)})$ measures the amount of local variation in the
estimate of $\vartheta_i$ contributed by the local variability of $j$th moment function at $P$.

Policy sensitivity
$$
S(\psi_W, \nu) 
= 
\int
\wt\psi_W 
\Big[
    \wt\nu - P \wt\nu \tfrac{dP}{dQ} / P  \tfrac{dP}{dQ}
\Big]
\tfrac{dP}{dQ}  \; dP
/
\int
\wt\nu
\Big[
    \wt\nu - P \wt\nu \tfrac{dP}{dQ} / P  \tfrac{dP}{dQ}
\Big]
\tfrac{dP}{dQ} \; dP
$$
gives the local counterfactual value $\psi_W(P) + h\cdot S(\psi_W,\nu) + o(h)$ of the economic
quantity $\vartheta$ identified with $\psi_W$ to the perturbation of size $h$ in the value of
statistical parameter $\nu(P)$ according to policy metric $L^2(Q)$. Measure $Q$ can be a policy
relevant reference distribution on the sample space. Taking the empirical measure $P$ as policy
measure is a convenient choice  in terms of estimation.

\section{Statistical surfaces} \label{sec:surfaces}

Let $\euP$ be a collection of probability measures on a sample space $(\calX,\calA)$.
The starting point for our investigation is to realize a statistical model as an object with
intrinsic geometry -- a space with notions of smoothness, length and angle.
In this section we consider a special case of a two-dimensional statistical model and
employ graphical aid to provide a nontechnical exposition.
The idea is to map a two-dimensional statistical model onto a surface in $\mR^3$ while preserving
the intrinsic metric properties of the model. We can then forget about the set of probability
measures and work with the surface in $\mR^3$. For details on geometry of surfaces we refer to
\cite{carmo1976differential}.

\vspace{.1cm}

The natural space to host statistical models is the set of square-integrable functions $L^2(\mu)$,
with some dominating measure $\mu$ for elements of the model $\euP$. In this ambient space,
probability distributions are identified with square-roots of their densities 
$dP^{1/2} \coloneqq \sqrt{\tfrac{dP}{d\mu}}$, the model $\euP$ is a subset of the unit ball of
$L^2(\mu)$, and the tangent set $T_P\euP$ is a subset of a hyperplane in $L^2(\mu)$. This simple
setup provides
a lot of structure to the model $\euP$, in particular, the information distance between two
distributions $P_0$ and $P_1$ is the length of the shortest curve on the model joining them. The
length of a curve $\a:[0,1]\ni t\mapsto P_t \in \euP$ is obtained by adding magnitudes of velocity
vectors along the curve:
\begin{align} \label{eqn:curveLength}
    L(\a) \coloneqq
    \int_{[0,1]} \sqrt{  \int_\calX \Big[ \tfrac{d}{dt} \;2\, dP_t^{1/2} \Big]^2 }  \; dt.
\end{align}
The curve in $L^2(\mu)$ is $t\mapsto dP^{1/2}_t$. Its velocity at time $t$ is the tangent vector
$v_t(x) = \tfrac{d}{dh}_{|h=t} dP_h^{1/2}(x)$, whose length, doubled for purely technical reasons,
$\norm{v_t}^2 = \int_\calX [2v(x)]^2 d\mu$ is the information metric norm. Finally,
the sum of velocities along the trajectory of the curve $\int_{[0,1]} \norm{v_t} dt$
is, by definition, the length of the curve.

\vspace{.1cm}

The problem of embedding $\euP$ into $\mR^3$ is to find a surface $S\subset\mR^3$ such that
length of the image of any curve $\a$ on $S$, computed according to the Euclidean geometry of
$\mR^3$, coincides with the value in \cref{eqn:curveLength}. 
Isometric embedding is an active area of research.  Conditions for preserving the metric are
formulated with a system of partial differential equations whose solvability requires enough
degrees of freedom provided by the dimensionality of ambient space. A general 2-manifold can be
embedded into $\mR^{10}$ by Nash's theorem and its extensions \cite{han2006isometric}.
The metric ultimately determines the shape of the surface required for the embedding.
\begin{assumptionRm}
Assume that $\euP$ is a smooth 2-manifold with metric given by \cref{eqn:curveLength} that
admits a smooth isometric embedding onto a regular surface $S\subset\mR^3$ at least locally at $P$.
\end{assumptionRm}
We consider three examples of statistical models with constant Gauss curvature:
\begin{figure}[h]
\centering
\begin{minipage}{0.31\textwidth}
\includegraphics[scale=0.5, trim=0in 7.0in 5in 0.5in, clip ]{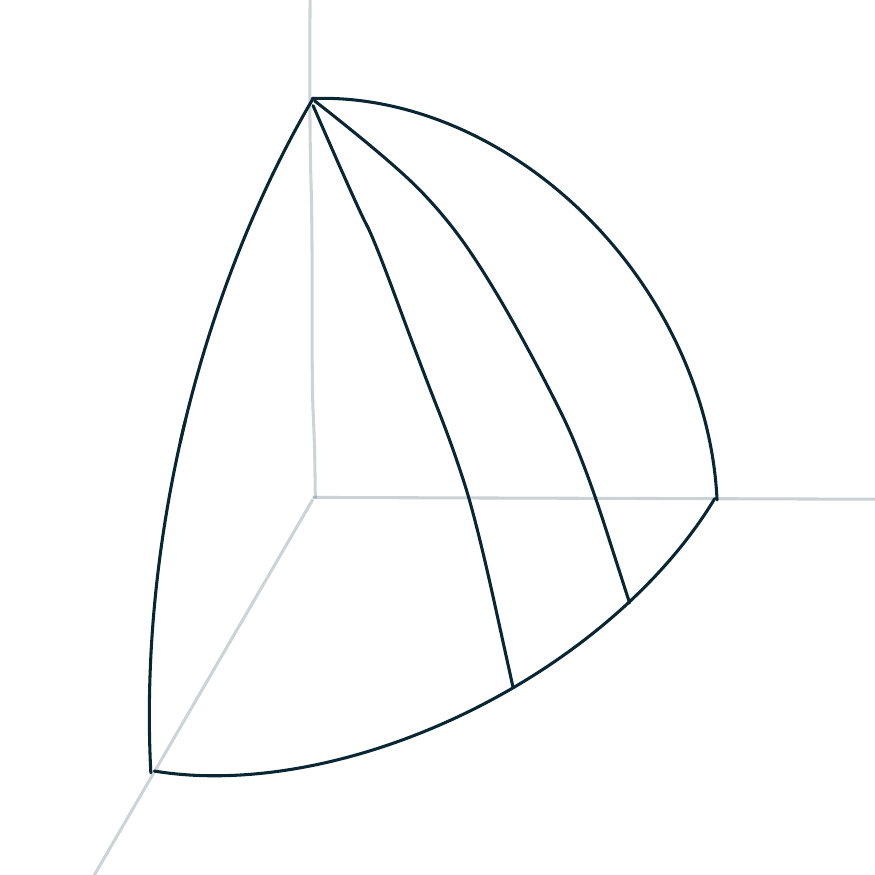}
    \subcaption{$K=1/4$}
    \label{fig:sphereMultinomial}
\end{minipage}
\begin{minipage}{0.31\textwidth}
\includegraphics[scale=0.5, trim=0in 7.5in 5in 0in, clip]{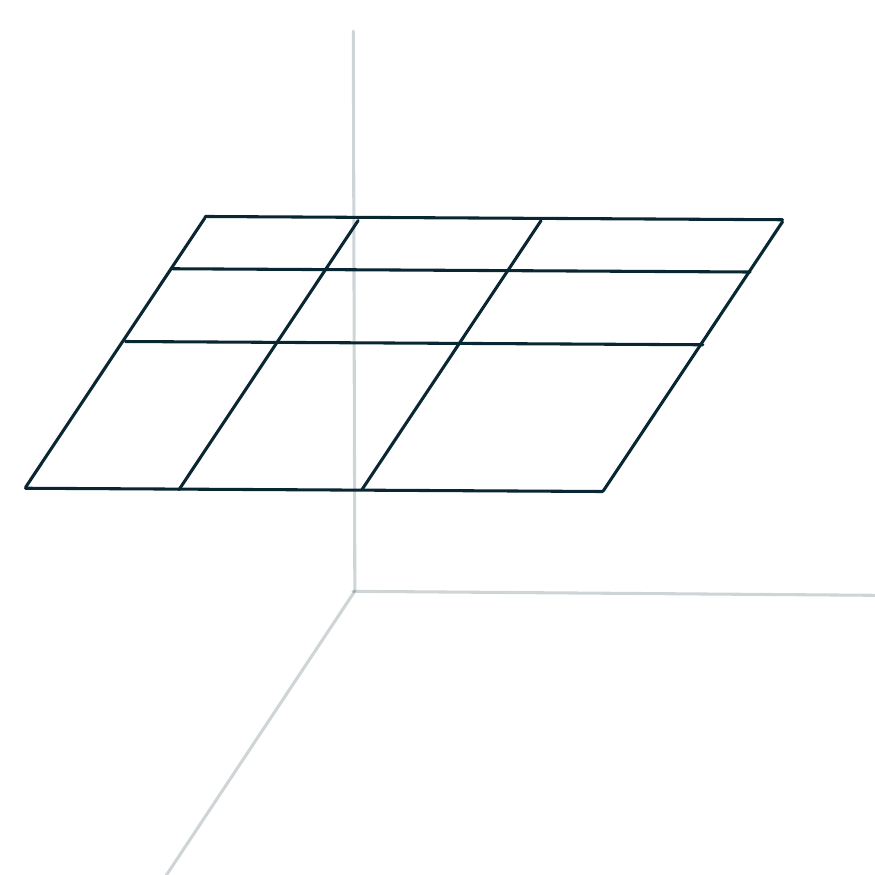}
    \subcaption{$K=0$}
    \label{fig:flatNormal}
\end{minipage}
\begin{minipage}{0.31\textwidth}
\includegraphics[scale=0.5, trim=1in 6.6in 4in 0.9in, clip]{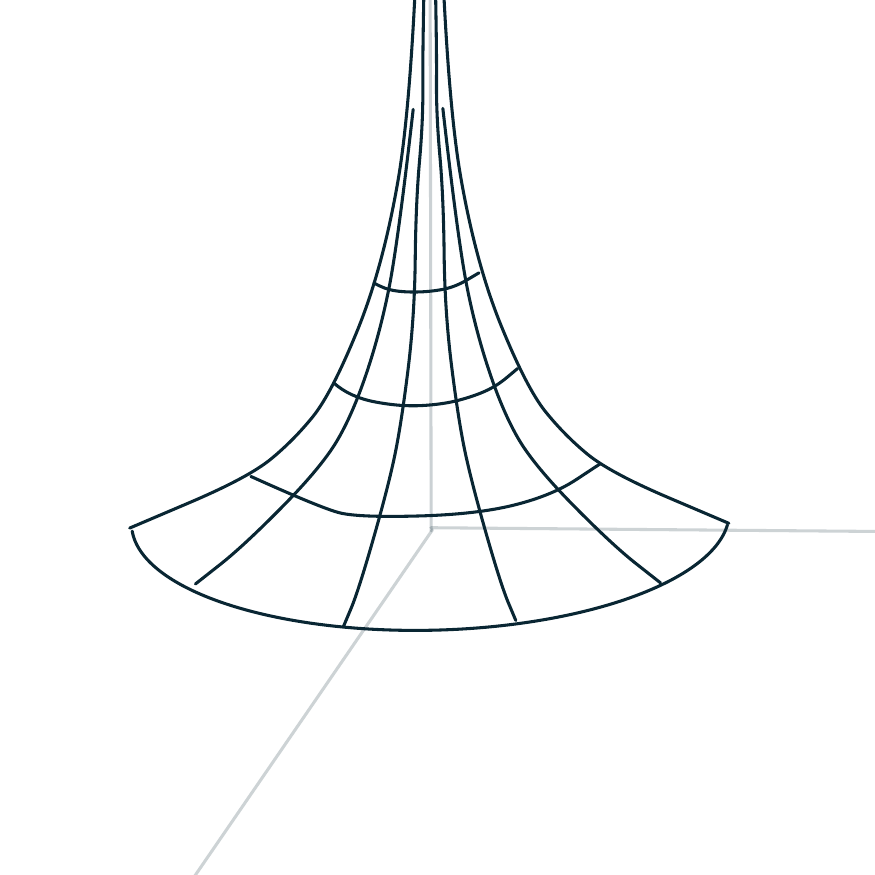}
    \subcaption{$K=-1/2$}
    \label{fig:hyperbolicNormal}
\end{minipage}
    \caption{2-dimensional statistical models with constant curvature}
    \label{fig:2dModels}
\end{figure}

\begin{example} \label{example:sphere}
    Multinomial family 
    $\euP_\mrm{sph} =\Set{\pi_1,\pi_2,\pi_3}{0\le\pi_i\le 1 \text{ and }
    \pi_1+\pi_2+\pi_3=1}$
    has Gauss curvature $1/4$ and isometric embedding
    onto an orthant of a sphere in $\mR^3$. See \cref{fig:sphereMultinomial}.
\end{example}

\begin{example} \label{example:flatNormal}
    Bivariate normal model 
    $\euP_\mrm{flat}
    =\Set{ N\big( (\mu_1,\mu_2), I_2 \big)}{\mu_1,\mu_2\in\mR}$
    with known variance
    has zero Gauss curvature and can be isometrically embedded into $\mR^3$ globally as a plane or
    locally onto a cylinder. See \cref{fig:flatNormal}.
\end{example}

\begin{example} \label{example:hyperbNormal}
    Univariate normal model 
    $\euP_\mrm{hyp}
    = \Set{ N(\mu,\s^2) }{\mu\in\mR, \s^2>0} $
    with location and scale parameters 
    has constant Gauss curvature $-1/2$.
    By Hilbert's theorem it has no global isometric imbedding into
    $\mR^3$ but is locally isometric to  the surface of a tractricoid (saddle shape).
    See \cref{fig:hyperbolicNormal}.
\end{example}
From a statistical model $\euP$ and its information metric we obtain a surface $S\subset\mR^3$
and from a statistical functional
$\psi(P)$ we obtain a function $f:S\into\mR$ defined on the points of the surface. We use
the surface to show that local behavior
of $f$ at $P\in S$, summarized by its derivative, determines the asymptotic behavior of estimates of
$\psi(P)$. Calculations near point $P$ on $S$ are carried out by means of a parametrization by an
open subset $U\subset\mR^2$.  There are many choices of a parametrization
$$
\mathbf{x}: \mR^2\supset U\ni(u,v) \mapsto \big(x(u,v), y(u,v), z(u,v)\big)\in S\subset\mR^3
$$
around a point $P$, the only requirements are that $\mbf{x}$ be differentiable with derivative
$d\mbf{x}_q:\mR^2\into\mR^3$ that is full rank for all $q\in U$.
For example, $\euP_{\mrm{sph}}$ can be parametrized by $x,y$ or $y,z$ or $z,x$ coordinates
of its points, or by latitude and longitude, or by points of the inscribed simplex.
Parametrization deforms a flat two-dimensional
neighborhood $U$ by stretching, shrinking and bending onto a neighborhood $V$ of the surface.
Because of the deformation, distances and angles in $U$ are different from those in $V$.
Calculations in each parametrization appear to be different but the values on the surface $S$ are
invariant similarly to how \textsc{mle} is parametrization invariant.

Differential calculus works on tangent vectors that are the infinitesimals.
At every point $P\in S$ there is a unique tangent plane $T_PS\subset\mR^3$ to the surface.
The derivative $d\mbf{x}_q$ of the parametrization maps vectors in $U$ anchored at $q$ into
tangent vectors in $T_{\mbf{x}(q)} S$.
Tangent vectors $\mbf{x}_u = d\mbf{x} e_1$ and $\mbf{x}_v =d\mbf{x} e_2$ span $T_{\mbf{x}(q)}S$
and are known as scores\footnote{I would appreciate a reference to the etymology of this terminology.}.
Due to deformation by $\mbf{x}$, orthonormal vectors $e_1,e_2$ in $U$ have images
$\mbf{x}_u, \mbf{x}_v$ that are not orthogonal and not unit length in $\mR^3$. This is because the
model $\euP$ is not flat at $P$ in its metric.  Consequently sensitivity $\cp_v u$ of parameters
$u,v$ on $S$ is not zero.
A function $f:S\into\mR$ is differentiable if its expression in local coordinates
$f\circ\mbf{x}$ is differentiable. The derivative $df_P:T_PS\into\mR$ maps tangent vectors to
$S$ at $P$ into vectors in $\mR$ anchored at $f(P)$.
\begin{figure}[h]
\centering
\label{fig:parmetrization}
\includegraphics[scale=0.7, trim= 0in 7.7in 0.5in 0.4in, clip=false, ]{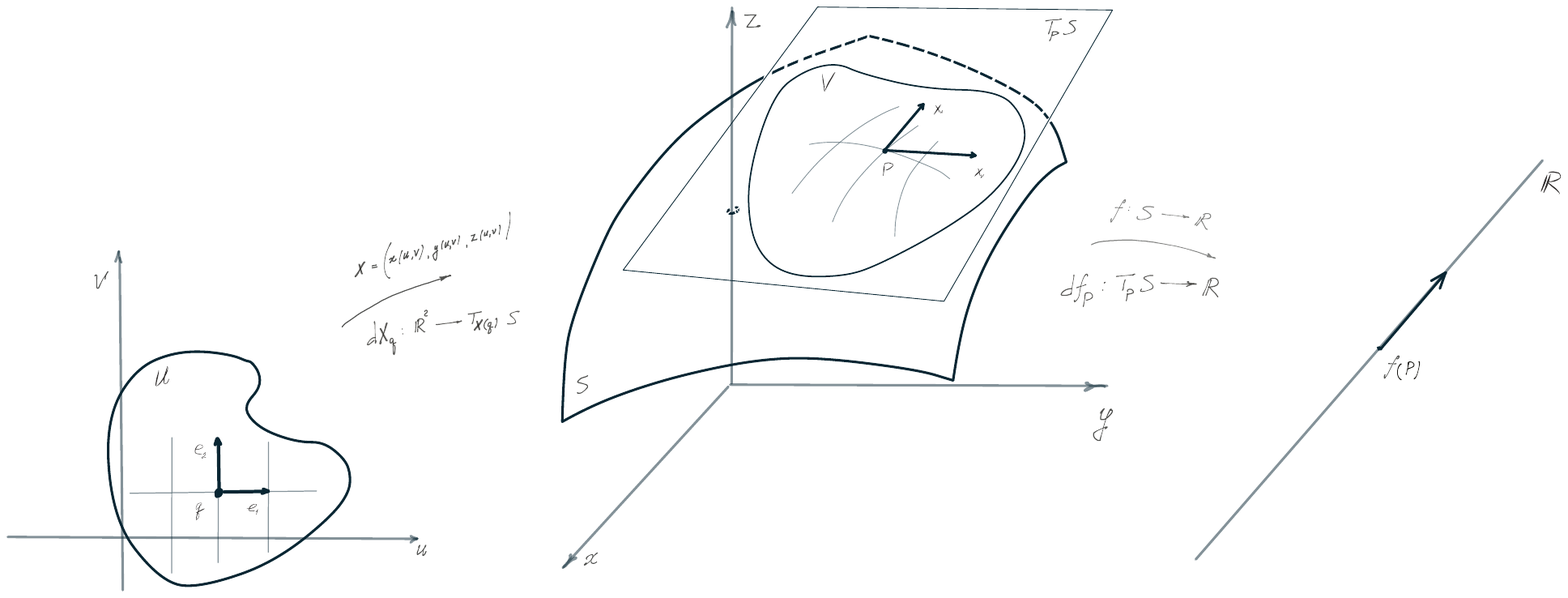}
\end{figure}

Recall that we took care to preserve distances and angles while mapping model $\euP$ into surface $S$.
The $\mR^3$ inner product $\brk{\,\cdot}{\cdot\,}_P$ induced on vectors of the tangent plane $T_PS$
is in agreement with intrinsic metric structure of the statistical model $\euP$.
This intrinsic statistical metric determines the sensitivity $\cp_\nu\psi$ of statistical
functional $\psi(P)$ to another parameter $\nu(P)$ as follows.
By a basic fact of linear algebra, the linear map $df_P$ has a simple representation by
the gradient vector $\nabla f_P$ of function $f$.
The gradient $\nabla f_P \in T_PS$ is the unique tangent vector that satisfies
\begin{IEEEeqnarray*}{c"t"c}
    \brk{\nabla f_P}{\mbf{x}_u}_{P} = df_P(\mbf{x}_u)
    & and &
    \brk{\nabla f_P}{\mbf{x}_v}_{P} = df_P(\mbf{x}_v).
\end{IEEEeqnarray*}
Gradient $\nabla f_P$ points in the direction on the surface along which values of
$f$ increase most rapidly and has magnitude $\norm{\nabla f_P}_{\mR^3}$ equal to the rate of the
increase at $P$ on model $\euP$. Since gradient $\nabla \nu_P$ determines the
linearization $w\mapsto\brk{\nabla \nu_P}{w}_P$ of functional $\nu$ at $P$, it is natural
to take it to be the ``$\nu$-direction" of the model at $P$. This is in perfect analogy with the
direction of $u$-axis in $U$ where the $u$ coordinate is the linear function
$w\mapsto\brk{e_1}{w}_{\mR^2}$ on $U$ with gradient $\nabla u = e_1$. This motivates our
measure of local statistical dependence:

%
\begin{defn} \label{def:sens2}
    Sensitivity of functional $\psi(P)$ to a statistical parameter $\nu(P)$ on statistical
    model $\euP$ is the directional derivative 
    \begin{IEEEeqnarray*}{c}
	\cp_\nu\psi(P) \coloneqq d\psi_P (\nabla\nu_P) = \brk{\nabla\psi_P}{\nabla\nu_P}_P.
    \end{IEEEeqnarray*}
    In the first equality we differentiate $\psi$ in the direction of $\nu$ given by the
    gradient of $\nu$.  Second equality follows from definition of the gradient of $\psi$.
\end{defn}

Next we use parametrization to compute the derivative $\cp_\nu\psi$ and establish
that parameter sensitivity of  \cref{def:sens2} and estimator sensitivity of
\cref{def:estSens} agree for many estimators, specifically that 
$\s_{\nu\nu}\Lambda(\wh\psi,\wh\nu) = \s_{\psi\nu}=\cp_\nu\psi$. Let
$    E(u_0,v_0) = \brk{ \mbf{x}_u}{ \mbf{x}_u }_P$, $F(u_0,v_0) = \brk{ \mbf{x}_u}{
\mbf{x}_v }_P$ and $G(u_0,v_0) = \brk{ \mbf{x}_v}{ \mbf{x}_v }_P$
%
%
denote the expression of the $\mR^3$ inner-product on $T_PS$ in local coordinates. And let 
$$ I_{u,v} =  
	\begin{bmatrix} 
	    E(u,v) & F(u,v) \\
	    F(u,v) & G(u,v)
	\end{bmatrix}
$$
denote the Fisher information matrix for this parametrization. Information matrix appears in the
expression for sensitivity because it reconciles distorted distances in $U$ with statistical
distances on $S$.
In local coordinates,
$f\circ\mbf{x}$ can be differentiated as usual to obtain partial derivatives $f_u,f_v$; these are
the directional derivatives of $f$ on $S$ along scores $\mbf{x}_u,\mbf{x}_v$. From relationships
$\brk{\nabla f}{ \mbf{x}_u } = f_u$ and $ \brk{\nabla f}{ \mbf{x}_v } = f_v$
%
%
we solve for the expression of $\nabla f$ in $\set{\mbf{x}_u,\mbf{x}_v}$ basis:
\begin{IEEEeqnarray}{c} \label{eq:gradFormula}
    \nabla f = \frac{G f_u - F f_v}{EG-F^2} \mbf{x}_u + \frac{E f_v - F f_u}{EG-F^2}\mbf{x}_v.
\end{IEEEeqnarray}

\begin{thm} \label{thm:sens2}
    Let $\euP$ be a two-dimensional statistical model with smooth isometric embedding $S$ into
    $\mR^3$. Let $\mbf{x}:U\into S$ be a parametrization of the model with information matrix
    $I_{u,v}$. Let $\psi,\nu$ be differentiable functionals defined on $\euP$.
    The directional derivative of $\psi(P)$ along $\nabla\nu$ is
    \begin{align} \label{eqn:cpInfluenceFcts}
    \cp_\nu\psi = \brk{\nabla\psi}{\nabla\nu} = 
    (I^{-1} [\psi_u \; \psi_v]^T)^T I (I^{-1} [\nu_u \; \nu_v]^T) = 
    [\psi_u \; \psi_v] I^{-1} [\nu_u \; \nu_v]^T.
    \end{align}
\end{thm}

\begin{corollaryRm} \label{thm:MLE2}
    In addition to conditions of \cref{thm:sens2} assume that for some function 
    $\dot\ell\in L^2(P_{u,v})$ and for every $(u_1,v_1)$ and $(u_2,v_2)$ in $U$
    $$
    (\log dP_{u_1,v_1}(x) - \log dP_{u_2,v_2}(x)) \le \dot\ell(x)\norm{(u_1,v_1)-(u_2,v_2)}
    $$
    and that \textsc{mle} estimators $(\wh{u},\wh{v})$ are consistent.
    Then \cref{eqn:asy} holds for \textsc{mle} plug-in estimators $(\wh\psi,\wh\nu)$,
    and parameter sensitivity $\cp_\nu\psi$ is equal to the estimator sensitivity
    $\s_{\psi\nu}$:
\begin{IEEEeqnarray}{c}
    \cp_\nu\psi = \sigma_{\psi\nu}.
\end{IEEEeqnarray}
\end{corollaryRm}

\begin{pf}
    Formula of \cref{thm:sens2} follows directly from isometry assumption and definition of
    gradient. Asymptotic normality \cref{eqn:asy} follows from \cite[p65, theorem
    5.39]{vaart2000asymptotic} by the delta method.
\end{pf}

\begin{figure}[h]
\centering
\includegraphics[scale=0.8, trim= 0in 6.7in 1.7in 1.0in, clip=true, ]{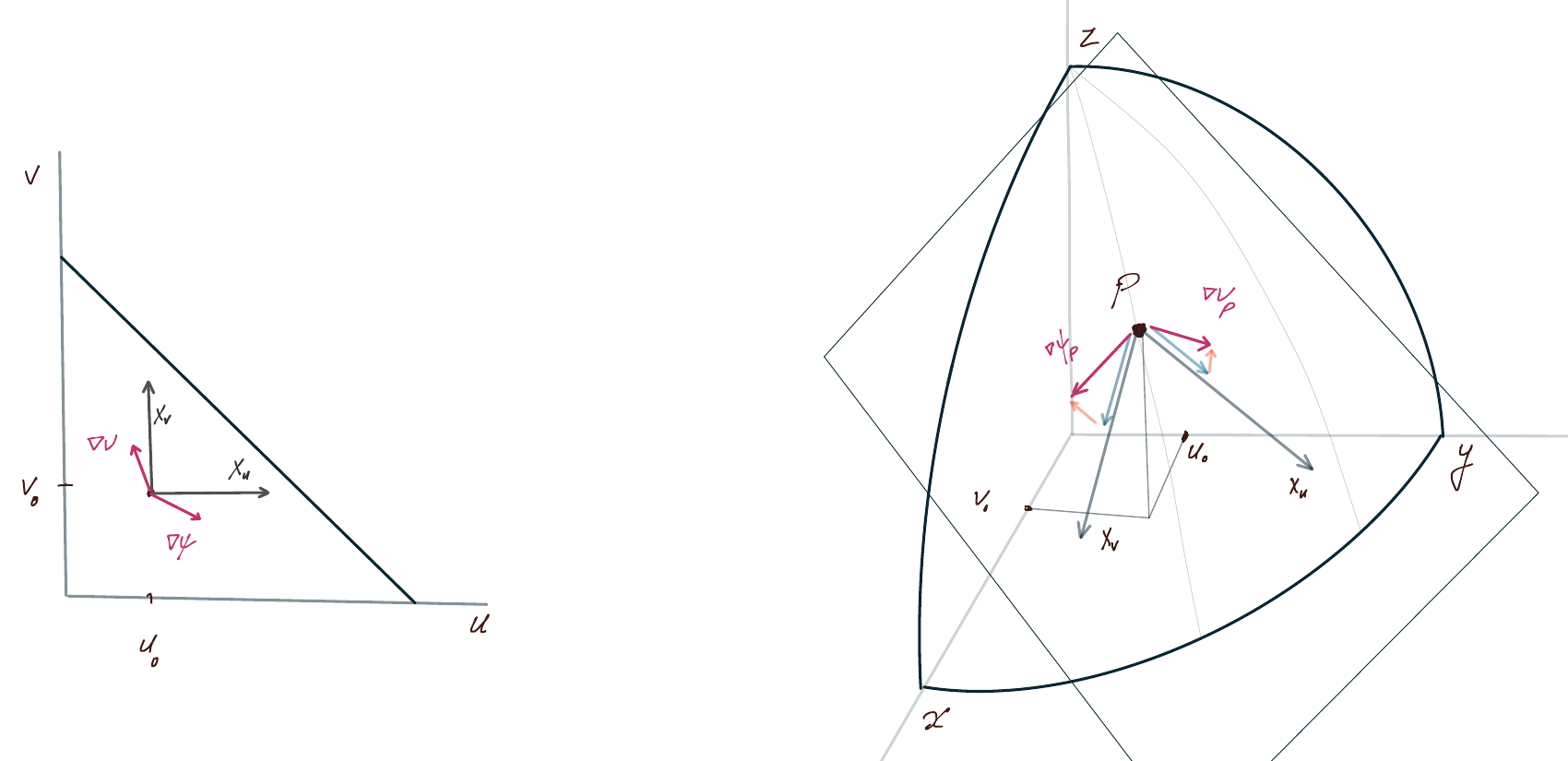}
\caption{}
\label{fig:sphereMultinomial1}
\end{figure}

\begin{example}[ \cref{example:sphere} continued]
    Using parametrization $\mbf{x}(u,v) = (2\sqrt{u}, 2\sqrt{v}, 2\sqrt{1-u-v})$, we compute the
    sensitivity of the functionals that make up the parametrization $\psi(u,v)=u$ and
    $\nu(u,v)=v$.
    The scores of the parametrization are
	$\mbf{x}_u=(u^{-1/2}, 0, -(1-u-v)^{-1/2})$ and 
	$\mbf{x}_v=(0,v^{-1/2}, -(1-u-v)^{-1/2})$.
    From \cref{eq:gradFormula} we compute 
    \begin{IEEEeqnarray*}{c"c}
	\nabla\psi = u(1-u) \mbf{x}_u -uv \mbf{x}_v &
	\nabla\nu = -uv\mbf{x}_u + v(1-v)\mbf{x}_v.
    \end{IEEEeqnarray*}
    Refer to \cref{fig:sphereMultinomial1}.  The sensitivity of the probability of first outcome to
    the probability of the second outcome is negative and decreases with each of the probabilities:
	$\cp_\nu\psi = -uv$.
\end{example}


\section{Geometry of statistical models} \label{sec:geom}

In this section we define general sensitivity measures of two statistical parameters.  
Sensitivity is defined through differential calculus of a statistical functional.
Functionals are real-valued maps of a set of possible distributions of each observation.
Sensitivity quantifies the local relationship between a functional of interest and any set of
regular functionals. We can relate this to regression and designate the functional of interest as
response or target and the set of regular functionals as controls. We only consider sensitivity to
a single control, but the extension to a set of controls is straightforward and the partialling
out reasoning of regression applies.  Sensitivity measures the deviation in the value of response
functional under the perturbation of the value of control functional.
Unlike regression coefficients, sensitivity is a bona fide directional derivative. The direction
depends on local properties of the control statistical functional and the notion of distance
between two distributions. Sensitivity can be used to make local counterfactual inferences about
economic quantities of interest in empirical work and to gain greater insight into asymptotic
distributions.

The set of possible sampling distributions is generally not linear, but can be modeled, in a
neighborhood of every point, as a smoothly transformed open subset of some linear space.
The idea takes some effort to develop methematically but the result provides great intuition.  
We introduce necessary elements of Riemannian geometry for completeness, and refer to 
do Carmo (1976, 1992) \cite{carmo1976differential, carmo1992riemannian} and  Lang (1999)
\cite{lang1999fundamentals} for more details.  
Most elements of differential geometry that we need to define sensitivity are also employed in the
semiparametric efficiency literature.
However, efficiency theory makes use of the ambient Hilbert space $H_2$ of square roots of
measures \cite[e.g.][p 739]{Koshevnik76}. From $H_2$ the model inherits the differential structure
(pathwise differentiability) and the information metric (Hellinger distance), similarly to our use
of $\mR^3$ in \Cref{sec:surfaces}.
By contrust, we define sensitivity based on a development of differential calculus on the model
without an ambient space and make dependence of sensitivity on the metric explicit. 
Our development is similar to the setup in van der Vaart (1991) \cite{vaart1991differentiable}.

The point here is to allow local counterfactual ``policy'' analysis at the population level to be
independent of the asymptotic approximations and statistical efficiency analyses.
We allow a general Riemannian metric on the model manifold, which we call a policy metric, to be
used for sensitivity measures at the population level.  We describe how these policy sensitivities
can be obtained from asymptotic distributions in \Cref{sec:sensRegEst}.

Let $M$ be a collection of distributions $P$ on sample space $(\calX,\calA)$.
We introduce a differentiable structure on $M$; this enables us to consider smooth functions
$\psi,\nu:M\into\mR$ which can be approximated on $M$ at a given point $P$ along directions $v\in T_PM$
of the tangent space;
differential $d\psi:T_PM\into\mR$ provides linear approximation of $\psi$ along any direction
$v$; metric $g$ is an inner-product on tangent spaces $T_PM$ that provides a Riesz representations
$\nabla\nu_P$ of the differentials $d\nu_P$ of $\nu$; finally, the sensitivity $\cp_\nu\psi$ is
the directional derivative $d\psi(\nabla\nu)$ of $\psi$ along the gradient direction of $\nu$.

We consider only the simplest case of an open manifold. Extensions that allow for manifolds with
boundaries, corners, etc., common with statistical models, are possible but are not considered here.
Tangent sets for the purposes of this paper are always complete linear spaces. Our approach to
start with an arbitrary manifold structure and consider inclusion into the space of square roots
of measures $H_2$ can be used to restrict the tangent space in an explicit way and allows us to
consider any metric in definition of sensitivity. 


\subsection{Differential structure.} \label{subsec:diff_str}

Statistical models can have many parametrizations. For example, the $N(\mu, I_3)$ family usually
parametrized
by the vector of means $\mu\in\mR^3$, can alternatively be specified using spherical coordinates 
$(\norm{\mu}, \tan^{-1}(\mu_2/\mu_1), \cos^{-1}(\mu_3/\norm{\mu})$; a (regression) function can be
parametrized by the coefficients of different Fourier bases. Parametrizations are necessary
for computation, but as long as we consider only compatible parametrizations, calculations we
do and quantities we define will be invariant of the chosen parametrization.
A differential structure is an equivalence class of compatible parametrizations. A manifold is a
set with a differentiable structure.

An atlas on $M$ is a collection of local parametrizations (charts)
$(U_i,\varphi_i)$ satisfying the following conditions:
\begin{description}[topsep=0pt, itemsep=0pt, partopsep=0pt, parsep=0pt,font=\normalfont,
    style=multiline]
    \item[AT1] Each $U_i$ is a subset of $M$ and the union of $U_i$ covers $M$.
    \item[AT2] Each $\varphi_i$ is a one-to-one and onto correspondence of $U_i$ with an open
	subset $\varphi_i(U_i)$ of a Banach space $E_i$ and for any $i,j$ $\varphi(U_i\cap U_j)$
	is open in $E_i$.
    \item[AT3] The composition 
	$\varphi_j \varphi^{-1}_i:\varphi_i(U_i\cap U_j)\into\varphi_j(U_i\cap U_j)$
	is a diffeomorphism for each pair of charts.
\end{description}

Let $M, N$  be manifolds. A map $f:M\into N$ is differentiable if, given $P\in M$, there are
charts $(U,\varphi)$ at $P$ and a chart $(V,\psi)$ at $f(P)$ such that $f(U)\subset V$ and the
composition 
$\psi \circ f \circ \varphi^{-1}:\varphi U\into\psi V$
is differentiable as a map between normed linear spaces. The composition is called expression of
$f$ in local coordinates.  Similarly, we define directional and compact differentiation 
\cite{0036-0279-22-6-R06,0036-0279-23-4-R02,penot2016analysis} by applying
the definition to the expression of $f$ in local coordinates.

\subsection{Tangent space.}

Let $E,F$ be Banach spaces. A tangent vector in $E$ is a direction
$v\in E$ with a position $P\in E$. Given a smooth curve $ [0,\e)\ni t \mapsto P_t \in E $
with position $P$ and direction $v=\frac{d}{dt} P_{t} \in E$ at time $t=0$, and a
differentiable map $f:E\into F$, we can associate the tangent vector $v$ with the directional
derivative operator
$$
\tfrac{d}{dt} f(P_t)_{\big |t=0} = Df_{P_t} \, (\tfrac{d}{dt}  P_t )_{\big| t=0} = D_v f (P).
$$
Let $M$ be a manifold modelled on a Banach space $E$. A curve on $M$ is a differentiable map
$\a:[0,\e)\into M$. A tangent vector at $\a(0)=P\in M$, corresponding to the direction of $\a$, is
the directional derivative operator $\a'(0)$ on differentiable maps $f:M\into\mR$
$$
\a'(0) f = \tfrac{d}{dt} (f \circ a)_{\big| t=0}.
$$
The set $T_PM$ of all tangent vectors to $M$ at point $P$, obtained from all curves passing
through $P$, is called the tangent space.
A tangent vector $\a'(0)$ corresponds to the direction $v\in E$ of the
expression $\varphi\circ\a_t$ of the curve in local coordinates. Tangent space $T_PM$ is in
bijective correspondence with $E$ and has the same structure of a topological vector space.
\begin{defn}
    Let $M,N$ be manifolds, let $f:M\into N$ be a differentiable map. For $P\in M$ and tangent
    vector $v\in T_P
    M$ let $\a_t$ be a curve with $\a_0=P$ and $\a'(0)=v$. Then $\b=f\circ\a$ is a curve in $N$.
    The \emph{differential} of $f$ at $P$ is the map
    $$df_P:T_P M \into T_{f(P)}N$$
    given by $df(v)=\b'(0)$. It is a continuous linear map between tangent spaces.
\end{defn}

\subsection{Metric.}

Geometric primitives discussed above are closely related to the ideas employed in semiparametric
efficiency. The next geometric primitive is implicit and fixed in the efficiency bounds theory but
has an active role in our local counterfactual analysis of functionals.  The idea is to give the
statistical model $M$ a notion of distance by giving each tangent space an inner product.  In
semiparametric efficiency theory this object is called information and it measures the
``statistical'' (Hellinger) distances between distributions.
However, the empirical researcher identifies statistical functionals $\psi,\nu:M\into\mR$ with
economic quantities $\vartheta,\eta$ and wants to understand the local relationship between
$\vartheta$ and $\eta$ at the data generating point $P$ on the model $M$. There is no reason to
assume that ``economic'' distances on $M$ coincide with ``statistical'' distances.
Therefore we consider a completely general metric for policy analysis purposes. 

A Riemannian metric on a statistical model $M$ is a correspondence $g$ that assigns to every point
$P\in M$ a continuous bilinear symmetric positive-definite form $g(\cdot,\cdot)_P$
on the tangent space $T_P M$, and varies smoothly over $M$.
For direction $v\in T_P M$ we can think of the norm $\abs{v}_g\coloneqq\sqrt{g(v,v)_P}$ as the
economic cost of a deviation from $P$ on $M$ at rate $v$. We will call $g$ a \emph{policy metric}
to contrast it with the statistical metric given by information  $\sqrt{ \int v^2 dP}$.  

\subsection{Sensitivity.}

The metric determines gradient directions of functions of $M$ as follows.
\begin{defn} \label{def:grad}
    Let $\psi:M\into\mR$ be a differentiable functional. The differential $d\psi_P:T_P M\into \mR$
    is a continuous linear map on the Hilbert space $(T_PM, g_P)$.  The Reisz representation
    vector $\nabla^g \psi_P \in T_PM$ of $d\psi_P$ is the \emph{gradient} of $\psi$ at $P$.
    It is the unique tangent vector that satisfies
    \begin{align*}
	d\psi_P(v) = g(\nabla^g \psi_P, v) \qquad \text{for every } v\in T_P M.
    \end{align*}
\end{defn}
\vspace{-16pt}
From definition it is clear that gradient of $\psi$ depends on the metric $g$.
The choice of metric determines the problem of approximating $\psi$ with a single tangent
vector.
By Cauchy-Schwarz,
$$
\sup_{\abs{v}\le 1} d\psi_P(v) \le \abs{\nabla^g\psi_P}.
$$
According to the metric $g$, gradient is the direction of most rapid increase
in the value of the function. 
The norm $\abs{\nabla \psi_P}_g$ is the slope of the tangent to the
restriction of $\psi$ along any curve through $P$ with unit speed and direction $\nabla \psi_P$.

\begin{defn} [(General sensitivity measures)]
    Let $\psi,\nu:M\into\mR$ be differentiable functionals on statistical model $M$ with policy
    metric $g$. Fix a point $P\in M$ on the model. The \emph{partial derivative} of $\psi$ with
    respect to $\nu$ or
%
%
\begin{align*}
    \text{the \emph{sensitivity} of $\psi$ to $\nu$ is}
    \hspace{7.7em}
    \cp_{\nu} \psi (P) \;\;
    &\coloneqq d\psi_P(\nabla\nu_P),
    \\
    \text{the \emph{sensitivity coefficient} of $\psi$ to $\nu$ is}
    \hspace{3em}
    S(\psi,\nu)_P &\coloneqq 
    \cp_\nu \psi / \abs{\nabla\nu}^2,
    \\
    \text{the \emph{local projection} of  $\psi$ onto $\nu$ at $P$ is}
    \hspace{2em}
    \Pi(\psi,\nu)_P &\coloneqq  
    S_{\nu}\psi \cdot \nabla\nu,
    \\
    \text{the \emph{local sufficiency} of  $\nu$ for $\psi$ at $P$ is}
    \hspace{2.4em}
    R(\psi,\nu)_P &\coloneqq 
    \abs{ \Pi_\nu\psi}^2 / \abs{\nabla\psi}^2.
\end{align*}
\end{defn}
\vspace{-16pt}

Clearly numbers $\cp_\nu\psi, S_\nu\psi,R_\nu\psi \in \mR$ and the linear map $\Pi_\nu\psi\in
T_PM^*$ depend on the choice of metric $g$ through gradients of $\psi,\nu$.
Directional derivatives $\cp_\nu\psi$ and $S_\nu\psi$ measure response in the value of
$\psi(P)$ to a perturbation in the value of $\nu(P)$ that is achieved by a deviation from $P$ on
$M$ in the direction of most rapid change in $\nu$.  This is analogous to partial derivatives in
linear spaces with respect to functionals of a coordinate system.
Projection vector $\Pi_\nu\psi$ gives the local approximation of $\psi$ by its partial derivative
along $\nu$ in all directions on $M$; this is the regression of $\psi$
onto $\nu$ locally at $P$. An interesting fact is that the coefficient of this local regression,
the sensitivity, is a genuine derivative in this case. 
Sufficiency is the coefficient of
determination in this regression and measures the alignment of $\psi(P)$ and $\nu(P)$ in a
neighborhood of $P$ on M. Specifically, $R_\nu\psi$ is the square of cosine of the angle between
$\nabla\psi$ and $\nabla\nu$.
A value of $R(\psi,\nu)$ close to $1$ reflects high degree of similarity in the local
behavior of $\psi,\nu$ at the data generating distribution; a value close to $0$ reflects that
$\psi,\nu$ move in orthogonal directions of the model $M$. When $R(\psi,\nu)$ is close to $1$ any
perturbation that moves $\nu$ will have a proportional effect on the value of $\psi$, where as
with $R(\psi,\nu)$ close to $0$ any perturbation that significantly moves $\nu$ will have
negligible effect on the value of $\psi$.

\begin{lemma} [Local counterfactual interpretation of sensitivity measures]
    \label{lem:interpSens}
    Suppose statistical model $M$ is a manifold. Researcher measures distances on $M$ with policy
    metric $g$ and is interested  in parameters $\psi,\nu$ that are smooth functionals
    $\psi,\nu:M\into \mR$ on the model. 
    Let $(-\e,\e)\ni t \mapsto P_t \in M$ be any smooth curve on $M$ with tangent vector
    $\tfrac{d}{dt}P_t = \nabla\nu_{P_0}$ at $t=0$. Then
    \begin{align*}
	\nu(P_t) = \nu(P_0) + t \cdot \abs{\nabla\nu}^2 + o(t)
	\quad\text{and}\quad
	\psi(P_t) = \psi(P_0) +  t\cdot\cp_\nu\psi + o(t)
	,
    \end{align*}
    so that sensitivity $S(\psi,\nu)$ is the local effect on $\psi$ of a change in the value of
    $\nu$ along $P_t$.

    Furthermore, the projection $\Pi_\nu\psi$ is the partial derivative (partial gradient) of
    $\psi$ along the gradient direction $\nabla\nu$ of parameter $\nu$: for any tangent vector
    $v\in T_P M$
    \begin{align*}
    t^{-1}[\psi(P+t v) - \psi(P)] 
    &=
    g(\Pi, v)
    +  g(\nabla\psi - \Pi, v) + o(1)
    \\
    &=
    S(\psi,\nu) \cdot d\nu[v] 
    + \mrm{residual}
    ,
    \end{align*}
    so that sensitivity $S(\psi,\nu)$ is the partial effect on $\psi$ from changing the value of $\nu$
    by \emph{any} local perturbation at $P$. The $R_\nu\psi$ measures the relative size of the
    partial derivative $\Pi_\nu\psi$ to total derivative $\nabla\psi$.

    Furthermore, the sufficiency $R(\psi,\nu) = \cos^2 \theta$, where the angle 
    \begin{align*}
	\theta =  \arccos 
	\Big(
	    g(\nabla\psi,\nabla\nu) / \abs{\nabla\psi} \abs{\nabla\nu}
	\Big)
    \end{align*}
    measures the alignment between $\psi$ and  $\nu$ locally at $P$.
\end{lemma}


Extension to a set of control functionals is straightforward by analogy with regression.
Here sensitivities are coefficients of the projection of $\nabla\psi$ onto the linear span of
$\nabla\nu_1,\ldots,\nabla\nu_p$. The interpretation of sensitivity coefficient of $\nu_1$ is as
above but for the local variation in $\psi$ and $\nu_1$ that is orthogonal to the linear span of 
$\nabla\nu_2\ldots\nabla\nu_p$ \cite{frisch1933,lovell1963}.

\section{Sensitivity of regular estimators} \label{sec:sensRegEst}

Let $M$ be a semiparametric model described in \Cref{sec:geom}, let $\psi,\nu:M\into\mR$
be Hadamard differentiable functionals on $M$. In this section we consider estimation based on
random samples from $P\in M$ and relate the asymptotic distribution of estimators
$\wh{\psi},\wh{\nu}$ to the sensitivity measures $\cp_\nu\psi, S_\nu\psi$ defined in
\Cref{sec:geom}.

Efficiency bounds on the asymptotic distribution of regular estimators $(\wh{\psi},\wh{\nu})$
depend on local properties of functionals $\psi,\nu$ on the
image $\euP$ of the inclusion 
$$
i:M\into{} H_2
$$
of the model manifold into the Hilbert space of square roots of measures, see Koshevnik and Levit
(1976) \cite{Koshevnik76} for the role of this embedding and Neveu (1965)
\cite[p. 112]{neveu1965mathematical} for the definition of the space $H_2$.\footnote{\cite{Koshevnik76}
    cite  \cite{neveu1965mathematical} but the English translation of \cite{Koshevnik76}
    references pages in the Russian translation of \cite{neveu1965mathematical}.}
We collect details of semiparametric efficiency theory in \Cref{sec:appendix}. 
Our setup with inclusion of $M$ into $H_2$ is similar to van der Vaart (1991)
\cite{vaart1991differentiable}, but we emphasise the intrinsic geometry of the model where as
\cite{vaart1991differentiable} is concerned with pathwise differentiability.
The following is a standard 

\begin{assumptionRm}
    Inclusion map $i:M\into H_2$ is differentiable with derivative $A_P$ that is a continuous
    linear map
    $$
    A:(T_PM,g) \into L^2_0(P)
    $$
    of tangent vectors $v$ to the model manifold $M$ into scores of parametric submodels that are
    $L^2(P)$ functions with mean zero $\int [Av](x)dP(x) =0$.
\end{assumptionRm}

Manifold $M$ determines the set of pathwise differentiable one-dimensional submodels $\euP(P)$ and
the tangent space $T_P\euP = A[T_PM] = R(A)\subset L^2_0(P)$, which are important elements of the
efficiency theory.
Note that differential $A$ need not be isomorphic and need not be isometric.
If range of $A$ is not closed in $L^2(P)$, then $A^{-1}$ is not bounded, and 
bilinear functional $g$ is not continuous on the tangent space $T_P\euP$.
For example, $T_PM=H^k$, the Sobolev space of $L^2(P)$ functions with $k$ derivatives.

We make the following stronger assumption that simplifies our functional analysis.
Roughly speaking, we consider models that behave either like finite dimensional smoothly
parametrized families or like fully nonparametric models.

\begin{assumptionRm} [(Regularity of policy geometry)] 
    \label{assume:inclReg}
    Inclusion map $i:M\into H_2$ is differentiable with derivative $A$ that is an isomorphism of
    $T_PM$ with $T_P\euP\subset L^2(P)$, in particular, $T_P\euP$ is closed and $A^{-1}$ is
    continuous.
\end{assumptionRm}

It follows that metric $g$ is continuous on the embedded tangent space
$(T_P\euP,\brk{\cdot\,}{\cdot\,}_P)$, and the $L^2(P)$ inner-product $\brk{\,\cdot}{\cdot\,}_P$
is continuous on the manifold tangent spce $(T_PM,g)$, and that inclusion differential $A$ has an
adjoint $A^*:T_P\euP\into T_PM$ such that
$$
\brk{Au}{v}_P = g(u, A^*v) \quad \text{for every } u\in T_PM, v\in T_P\euP.
$$
Since $A$ is the derivative of the inclusion map, we can treat it as the identity operator on
$T_PM$. Furthermore, from functional analysis identity $(\Ker A)^\perp = \widebar{\Ran A^*}$ and
continuity of $A^{-1}$, conclude that $A^*$ has a continuous inverse $(A^*)^{-1}:T_P\euP\into
T_PM$, and
\begin{align} \label{eq:adjoint}
g(u,v) = \brk{u}{ (A^*)^{-1} v}_P \quad \text{for every } u,v\in T_PM.
\end{align}

\begin{thm}[Relationship between sensitivity and efficiency]
    Let $(M,g)$ be a statistical model with a policy metric as in \Cref{sec:geom}.
    Let $\psi,\nu:M\into\mR$ be Hadamard differentiable with gradients $\nabla\psi,\nabla\nu$;
    suppose policy regularity \Cref{assume:inclReg} holds; then functionals
    $\psi,\nu:\euP\into\mR$ are differentiable with influence functions $\wt{\psi},\wt{\nu}$ and
    \begin{align*}
    \cp_\nu\psi = \brk{\wt{\psi}}{A^*\wt{\nu}}_P \quad\text{and}\quad
    S_\nu{\psi} = \brk{\wt{\psi}}{A^*\wt{\nu}}_P / \brk{\wt{\nu}}{A^*\wt{\nu}}_P.
    \end{align*}
\end{thm}
\begin{pf}
    Differentiability of $\psi,\nu:\euP\into\mR$ follows from \cite{vaart1991differentiable} or
    directly from $i$ being a diffeomorphism by the inverse function theorem. From
    \cref{eq:adjoint} and \cref{def:grad} we have
    \begin{align} \label{eq:gradOp}
    \wt{\psi}  = (A^*)^{-1} (\nabla^g\psi) \quad\text{and}\quad
    \nabla^g{\psi}   = A^* (\wt\nu).
    \end{align}
    Formulas for policy directional derivative and sensitivity follow from \cref{eq:adjoint} by
    substituting above expression for $\nabla\psi$ and the same expression for $\nabla\nu$.
\end{pf}

\begin{defn} \label{def:gradOp}
We will call $A^*$ the gradient operator because of \cref{eq:gradOp}.
\end{defn}

\begin{corollaryRm}[(Characterization of estimator sensitivity)]
    Let $\wh{\psi}_n,\wh{\nu}_n$ be regular efficient estimators of $\psi,\nu$ on statistical
    model $M$.
    Then estimator sensitivity $\Lambda(\wh\psi,\wh\nu)$ is the sensitivity $S_I(\psi,\nu)$ with
    respect to the information metric on $M$.
    Estimator sufficiency $\Delta(\wh\psi,\wh\nu)$ is the information sufficiency of $\nu$ for
    $\psi$ and, in addition to interpretations of \Cref{lem:interpSens}, is the efficiency gain in
    estimation of $\psi(P)$, obtained by restricting the statistical model by setting the value of
    $\nu$ to its population value $\nu(P)$.
\end{corollaryRm}
\begin{pf}
    From the nonparametric version of H\'{a}jek's convolution theorem (see \Cref{convolution}),
    the asymptotic distribution of a regular estimator
    $(\wh{\psi}_n,\wh{\nu}_n)$ is $N(0,\Sigma)* \calN$, where 
    $\Sigma_{\psi\psi} = \norm{\wt{\psi}}_P^2$,
    $\Sigma_{\psi\nu} = \brk{\wt\psi}{\wt\nu}_P$ and 
    $\Sigma_{\nu\nu} = \norm{\wt{\nu}}_P^2$. 
    For any efficient estimator sequence, the noise term
    $\calN$ is a point mass at zero so that asymptotic distribution is just $N(0,\Sigma)$. If
    metric $g$ on $M$ is given by the information inner-product of $L^2(P)$ at $P$, then operator
    $A$ is a unitary isometry, $A^*$ is the identity operator on $T_P\euP$. It follows that
    influence functions $\wt\psi,\wt\nu$ are gradients $\nabla^I\psi,\nabla^I\nu$ and
    sensitivity is the asymptotic covariance $\cp_{\nu}\psi = \brk{\wt\psi}{\wt\nu}_P$.
    Characterization of estimator sufficiency follows from considering the restricted model
    $M_\nu$ that is a local submanifold of $M$ determined by the closed subspace 
    $$T_PM_\nu = \Set{v\in T_PM}{\brk{v}{\wt{\nu}}_P=0}$$
    of the tangent space $T_PM$ (see \cite[ch2 \S2]{lang1999fundamentals}).
    Efficient influence function on $M_\nu$ is
    $$\nabla^I_{M_\nu}\psi = \nabla^I_{M}\psi - \Pi(\psi,\nu),$$
    therefore estimator sufficiency $\Lambda(\psi,\nu)$ gives the reduction in the
    asymptotic variance of a regular efficient estimator of $\psi$ on $M_\nu$ relative to the
    bound for $M$.
\end{pf}

Estimator sufficiency is informative of the local statistical relationship between $\psi$ and
$\nu$, and specifically, to what extent regular estimates of parameter $\nu(P)$ determine
inferences based on the asymptotic distribution of regular estimators of $\psi(P)$ in the sense of
efficiency gain.

\subsection{Gradient operator of an absolutely continuous policy measure.} \label{ex:mult}
%

Here we consider a tractable example of policy and obtain explicit relationships between policy
gradients and influence functions.
Consider a nonparametric model $M$, fix a distribution $P\in M$, and let the tangent space 
$T_P M=T_P\euP=L^2_0(P)$ be unrestricted. Suppose that policy metric $g$ on $T_PM$ is given by 
\begin{align} \label{eqn:PolicyMetric}
    g (u,v) = \int u(x) v(x) dQ(x) \qquad u,v\in L^2_0(P)
\end{align}
where policy distribution $Q$ satisfies the following regularity condition
\begin{assumptionRm}
    \label{assume:AbsCont}
    $Q\ll P$ and the likelihood ratio satisfies $0 < m \leq \tfrac{dQ}{dP}(x) \leq
    M < \infty$.
\end{assumptionRm}

Probability measure $Q$ may be a social weighting on sample space $\calX$ that is relevant for
policy. Policy probability density $dQ(x)$ is the cost of displacing a unit of mass in $P$ at
location $x$ of the sample space $\calX$.

We want to find the policy relevant response $\cp_\nu \psi = g(\nabla\psi, \nabla\nu)$ of the
change to economic quantity associated with statistical functional $\psi$ that would result from a
perturbation to $\nu$.
This can be computed from influence functions, obtained as part of the asymptotic distribution
derivation for estimators of $\psi,\nu$ or from an efficiency bound calculation.  We assume that
policy regularity condition \cref{assume:inclReg} and find the gradient operator $A^*$, which we
can then verify to be isomorphic.

From definition \cref{def:grad} we have the following relationships 
\begin{align*}
     d\psi_P (v) &= g(\nabla\psi, v) 
		  = \brk{\wt\psi}{v}_P,
		  \quad v\in T_PM, T_P\euP.
\end{align*}
It follows that for every $v\in L^2_0(P)$
\begin{align*}
    \int \wt{\psi} \, v  \,dP 
    &=
    \int \nabla\psi \tfrac{dQ}{dP} \, v \,  dP
    \\
    &=
    \int 
    \Big[
	\nabla\psi \tfrac{dQ}{dP} - P[ \nabla\psi \tfrac{dQ}{dP} ]
    \Big]
    \, v \,  dP
\end{align*}
so that 
$$
\wt{\psi} =  \nabla\psi \tfrac{dQ}{dP} - P [\nabla\psi \tfrac{dQ}{dP}]
\quad \text{ and } \quad
\nabla \psi = \Big[\wt\psi +  P[\nabla\psi \tfrac{dQ}{dP}] \Big]\tfrac{dP}{dQ}
\qquad \text{ a.e. } P,Q.
$$
To solve for the centering constant $P[\nabla\psi \tfrac{dQ}{dP}]$, use the fact that 
$P\nabla\psi = 0$, to find that 
$P[\nabla\psi \tfrac{dQ}{dP}] = - P\wt\psi\tfrac{dP}{dQ} / P \tfrac{dP}{dQ}$. Conclude:
\begin{align}
    \label{eqn:policyGrad}
    \nabla \psi 
    =
    \Big[\wt\psi -  P\wt\psi\tfrac{dP}{dQ} / P \tfrac{dP}{dQ}  \Big]\tfrac{dP}{dQ}.
\end{align}
Thus, we have expressed the policy gradients $\nabla\psi,\nabla\nu$ in terms of the influence functions
and can compute the policy sensitivity as follows.

\begin{thm} [Policy measure sensitivity]
Suppose that policy metric given by \cref{eqn:PolicyMetric} satisfies \cref{assume:AbsCont}.
Then \cref{assume:inclReg} holds and policy sensitivity of statistical functionals $\psi,\nu$
is
\begin{align*}
    \cp_\nu \psi &= g(\nabla\psi,\nabla\nu) 
     \\
     &= 
     \int 
     \Big[\wt\psi -  P\wt\psi\tfrac{dP}{dQ} / P \tfrac{dP}{dQ}  \Big]\tfrac{dP}{dQ}
     \, 
     \Big[\wt\nu -  P\wt\nu\tfrac{dP}{dQ} / P \tfrac{dP}{dQ}  \Big]\tfrac{dP}{dQ}
     \,  dQ 
     \\
     &=
     \int \wt\psi \,
     \Big[\wt\nu -  P\wt\nu\tfrac{dP}{dQ} / P \tfrac{dP}{dQ}  \Big]\tfrac{dP}{dQ}
     \, dP.
\end{align*}
And the gradient operator is the multiplication operator:
\begin{align} \label{multGradOp}
    A^* v = \Big[v - P v\tfrac{dP}{dQ} / P \tfrac{dP}{dQ} \Big] \tfrac{dP}{dQ} 
    \quad \text{and} \quad
    (A^*)^{-1} u = u\tfrac{dQ}{dP} - Qu,
    \quad 
    v\in T_P\euP,
    u\in T_PM.
\end{align}
\end{thm}

\begin{remark}
    This is similar to propensity score reweighting. The likelihood ration $\tfrac{dP}{dQ}$
    adjusts for the discrepancy between the policy distribution and sampling distribution.
\end{remark}

\begin{remark}
    The condition $P\ll Q$ is not necessary if $\tfrac{dP}{dQ}$ is understood to be the density of
    the absolutely continuous part of $P$ in the Lebesgue decomposition with respect to $Q$.
\end{remark}

\subsection{Estimating sensitivity.} 
%

Reporting sensitivity in empirical work requires 
estimating it along with the asymptotic variance. We consider two distinct scenarios. If the
policy metric $g_P$ has a fixed relationship with the distribution $P$ of data, then estimating
sensitivity is straightforward
and consistency follows (roughly) from consistency of the asymptotic approximation. If the policy
metric $g_P$ depends on the distribution $P$ in a general way, then gradient operator $A^*_P$ needs
to be estimated and consistency requires additional justification.

We consider the typical situation where one estimates a vector of parameters $\th\in\Theta$ and
obtains an estimate of the asymptotic variance by plugging in the estimate $\wh\th$ to obtain
influence functions $\wt\psi_{\wh\th},\wt\nu_{\wh\th}$. Here $\psi,\nu$ could be some functions of
$\th$.
We first assume that $g_P$ has a fixed relationship to $P$ so that the gradient operator $A^*$ is
known. We use notation $\mP_n = n^{-1}\sum_{i=1}^n\d_{X_i}$ for the empirical measure.

\begin{thm} \label{thm:sensConsistency1}
    Let $\calF=\Set{\wt\psi_\th \cdot A^* \wt\nu_\th}{ \th\in\Theta}$ be a Glivenko-Cantelli class
    of functions; let $\wt\psi_{\wh\th(n)}\into \wt\psi$ and $\wt\nu_{\wh\th(n)}\into \wt\nu$ in
    $L^2(P)$. Then 
    $$\widehat{\cp_\nu\psi} = \brk{\wt\psi_{\wh\th(n)}}{A^*\wt\nu_{\wh\th(n)}}_{\mP_n}$$ 
    is a consistent estimator of sensitivity $\cp_\nu\psi$.
\end{thm}
\begin{pf}
    By triangle inequality
\begin{multline*}
    | \brk{\wt\psi_{\wh\th(n)}}{A^*\wt\nu_{\wh\th(n)}}_{\mP_n} -
    \brk{\wt\psi}{A^*\wt\nu}_{P} | \\
    \le
    \underbracket[0.2pt][2pt]{
    | \brk{\wt\psi_{\wh\th(n)}}{A^*\wt\nu_{\wh\th(n)}}_{\mP_n} -
    \brk{\wt\psi_{\wh\th(n)}}{A^*\wt\nu_{\wh\th(n)}}_{P} |
    }_{I} +
    \underbracket[0.2pt][2pt]{
    | \brk{\wt\psi_{\wh\th(n)}}{A^*\wt\nu_{\wh\th(n)}}_{P} -
    \brk{\wt\psi}{A^*\wt\nu}_{P} |
    }_{II}
\end{multline*}
We use uniform law of large numbers over the class $\calF$ to control term $I$
\begin{align*}
    I \le
    \sup_\th
    | \brk{\wt\psi_{\th}}{A^*\wt\nu_{\th}}_{\mP_n} -
    \brk{\wt\psi_{\th}}{A^*\wt\nu_{\th}}_{P} |.
\end{align*}
We use triangle inequality, Cauchy-Schwarz and $L^2(P)$ convergence to control term $II$
\begin{align*}
    II 
    &\le
    | \brk{\wt\psi_{\wh\th(n)}}{A^*\wt\nu_{\wh\th(n)}}_{P} -
    \brk{\wt\psi_{\wh\th(n)}}{A^*\wt\nu}_{P} | + 
    | \brk{\wt\psi_{\wh\th(n)}}{A^*\wt\nu}_{P} -
    \brk{\wt\psi}{A^*\wt\nu}_{P} | \\
    &\le
    \norm{\wt\psi_{\wh\th(n)}}_P \norm{A^*} \norm{\wt\nu_{\wh\th(n)} - \wt\nu}_P +
    \norm{\wt\psi_{\wh\th(n)} - \wt\psi}_P \norm{A^*}  \norm{\wt\nu}_P.
\end{align*}
\end{pf}

With the additional assumption that functions $\Set{\wt\nu_\th\cdot A^*\wt\nu_\th}{\th\in\Theta}$ are
Glivenko-Cantelli, one can form a consistent estimator of sensitivity coefficient $S_\nu\psi$.
More primitive conditions can be based on e.g. bracketing entropy. If an estimate
of bracketing numbers is available for functions $\wt\nu_\th$ and $A^*$ preserves point-wise order
at each $x\in\calX$ like the multiplication operator \Cref{ex:mult}, then one can estimate
bracketing numbers for $A^*\wt\nu_\th$.
%
%

Estimator of sensitivity derivative when gradient operator $A^*_P$ depends on $P$ can be based on
the plugin estimate with empirical distribution or mollified empirical distribution
\begin{align} \label{eqn:genDerEst}
    \widehat{\cp_\nu\psi} = \brk{\wt\psi_{\wh\th(n)}}{A^*_{\wh\mP(n)}\wt\nu_{\wh\th(n)}}_{\mP_n}.
\end{align}
E.g. the multiplication operator of \Cref{ex:mult} is of this form because the likelihood ratio
$\frac{dP}{dQ}$ of the data generating $P$ to policy cost distribution $Q$ depends on unknown $P$.
We leave consistency of the general form \cref{eqn:genDerEst} to future work and consider
consistency of policy sensitivity of \Cref{ex:mult} formulated with a policy cost distribution
$Q$.

\begin{thm} [Consistency of plug-in estimator of policy measure sensitivity]
Suppose (i) functions 
    $$
    \Set{\wt\psi_\th^2\wt\nu_\th^2, \wt\psi_\th\wt\nu_\th\tfrac{dP}{dQ}, \wt\psi^2_\th,
	\wt\nu_\th^2, \wt\psi_\th \tfrac{dP}{dQ}, \wt\nu_\th \tfrac{dP}{dQ} }{ \th\in\Theta}
    $$
    are Glivenko-Cantelli;
(ii)
    $\wt\psi_{\wh\th} \into \wt\psi_{\th}$ and $\wt\nu_{\wh\th} \into \wt\nu_{\th}$ in $L^4(P)$;
(iii)
    estimator of likelihood ratio is consistent in the empirical MISE sense
    $\mP_n\Big[\widehat{\tfrac{dP}{dQ}} - \tfrac{dP}{dQ}\Big]^2 \into 0$ in $P$;
(iv)
    likelihood ratio $\tfrac{dP}{dQ}$ satisfies \Cref{assume:AbsCont}.
Then plug-in estimator of policy sensitivity
\begin{align} \label{eqn:costPolicyDer}
    \widehat{\cp_\nu\psi} 
    =
    \mP_n 
    \Big\{
    \wt\psi_{\wh\th} \;
    \Big[
    \wt\nu_{\wh\th} 
    - 
    \mP_n \wt\nu_{\wh\th} \widehat{\tfrac{dP}{dQ}} / \mP_n\widehat{\tfrac{dP}{dQ}}
    \Big] \;
    \widehat{\tfrac{dP}{dQ}}
    \Big\}
\end{align}
is consistent.
\end{thm}

\begin{pf}

We formulated conditions for influence functions and likelihood ratio estimator independently.
Our strategy is to avoid interacting influence functions with likelihood ratio estimates in terms
that require uniform convergence. This is accomplished with Cauchy-Schwarz and triangle
inequalities. Let $r=\tfrac{dP}{dQ}$ and $\wh r=\widehat{\tfrac{dP}{dQ}}$. We need to control
convergence of the following two remainder terms:
\begin{align*}
    \abs{\widehat{\cp_\nu\psi} - \cp_\nu\psi}
    \le
    \underbracket[0.2pt][2pt]{
    \abs[\Big]{
	\mP_n \wt\psi_{\wh\th} \wt\nu_{\wh\th} \wh r -  P \wt\psi_\th \wt\nu_\th r
    } 
    }_{I}
    +
    \underbracket[0.2pt][2pt]{
    \abs[\Big]{
	\mP_n \big[\wt\nu_{\wh\th} \wh r \big] \; \mP_n \big[\wt\psi_{\wh\th} \wh r\big] / \mP_n \wh r    
	-
	P \big[\wt\nu_{\th} r\big] \; P \big[\wt\psi_{\th} r\big] / P r    
    }
    }_{II} .
\end{align*}
To separate $\wt\psi_{\wh\th}\wt\nu_{\wh\th}$ from $\wh r$ in term $I$ we center at 
$\mP_n \wt\psi_{\wh\th}\wt\nu_{\wh\th} r$ and use triangle inequality to obtain terms $Ia$ and $Ib$:
\begin{align*}
Ia = \abs { \mP_n \wt\psi_{\wh\th} \wt\nu_{\wh\th} \wh r - \mP_n \wt\psi_{\wh\th} \wt\nu_{\wh\th} r }
\le
\sqrt{\mP_n [\wt\psi_{\wh\th} \wt\nu_{\wh\th} ]^2} \sqrt{\mP_n [\wh r - r]^2}.
\end{align*}
The first term on the right of $Ia$ is $O_p(1)$ by the uniform and $L^4$ convergences, where as the
second term is $o_P(1)$ by assumption (iii) on the estimator of likelihood ratio. Term
\begin{align*}
Ib = \abs { \mP_n \wt\psi_{\wh\th} \wt\nu_{\wh\th} r - P \wt\psi_\th \wt\nu_\th r }
\le
\abs { \mP_n \wt\psi_{\wh\th} \wt\nu_{\wh\th} r - P \wt\psi_{\wh\th} \wt\nu_{\wh\th} r } +
 P \abs { \wt\psi_{\wh\th} \wt\nu_{\wh\th} - \wt\psi_\th \wt\nu_\th } r
\end{align*}
is $o_P(1)$ by the assumed uniform convergence, uniform bound on likelihood ratio and
$L^4$ convergence of influence functions with plug-in.

We center term $II$ at $\mP_n \wt\nu_{\wh\th}\wh r P\wt\psi r / \mP_n\wh r$ and use triangle
inequality to obtain terms $IIa,IIb$:
\begin{align*}
    IIa =
    \underbracket[0.2pt][2pt]{
    \abs[\big]{\mP_n \wt\nu_{\wh\th}\wh r / \mP_n\wh r}
    }_{IIa3/IIa4}
    \Big[
    \underbracket[0.2pt][2pt]{
	\abs{\mP_n \wt\psi_{\wh\th} \wh r - \mP_n \wt\psi_{\wh\th} r}
    }_{IIa1} +
    \underbracket[0.2pt][2pt]{
	\abs{\mP_n \wt\psi_{\wh\th} r - P \wt\psi_\th r }
    }_{IIa2}
    \Big] .
\end{align*}
Term $IIa1$ is controlled similarly to term $Ia$ and term $IIa2$ similarly to term $Ib$. Term
$IIa3$ is bounded by a Cauchy-Schwarz estimate. Term $IIa4$ is bounded by $(iii)$ and law of large
numbers for $\mP_n r$.
\begin{align*}
    IIb =
    \abs{\mP_n \wt\nu_{\wh\th} \wh r / \mP_n \wh r - P \wt\nu_{\th} r / P r} \abs{ P\wt\psi r} ;
\end{align*}
Term $\mP_n \wt\nu_{\wh\th} \wh r$ converges to $P \wt\nu_{\th} r$ by the argument for term $I$.
From (iii), Cauchy-Schwarz and law of large numbers have
$\abs{\mP_n \wh r - P r} \le \sqrt{\mP_n [\wh r - r]^2} + \abs{\mP_n r - P r} = o_P(1)$. Conclude
that $IIb$ is $o_P(1)$ by continuous mapping argument.
\end{pf}

Possible variation on above strategy is to assume that likelihood estimates are bounded and apply
H\"{o}lder's inequality instead of Cauchy-Schwarz.
An alternative strategy is to investigate uniform convergence of the product of influence
functions and likelihood ratio approximations.

\section{Examples} \label{sec:examples}

Here we continue with our example setup of a nonparametric model $M$ with full tangent space
$T_PM=T_P\euP=L^2_0(P)$ on sample space $\calX=\mR$ with Borel $\s$-algebra.
Ichimura and Newey (2015) \cite[IN][]{ichimura2015influence} describe how
influence functions can be computed. Their idea is to use Lebesgue differentiation to recover the
influence function $\wt\psi\in L^2(P)$ from its integral in \cref{def:grad}.
It is enough to consider a sequence of curves 
$P_{z,t}^j = (1-t)P + tG_z^j$ and compute 
\begin{align*}
    \wt\psi(z)=\lim_j \left[ \tfrac{d}{dt} \psi(P_{z,t}^j)_{\big|t=0} \right]
\end{align*}
for
an approximation to identity $G_z^j\into \d_z$. In models with tangent sets that are a proper
subspaces of $L^2_0(P)$, the efficient influence function is the projection onto the subspace.

\subsection{Mean}
%
functional $\psi_1(P) = \int_\mR x \, dP(x)$ has information gradient
$\wt\psi_1(x) = x - \psi(P) \in L^2_0(P)$.
\begin{align*}
    \tfrac{d}{dt}\psi(P_{z,t}^j) = \tfrac{d}{dt} \int x f_t(x) \, dx
= \tfrac{d}{dt} \int x \Big[ f(x) + t\{g_z^j(x)-f(x) \} \Big] \, dx  \\
= \int x \Big[ g_z^j(x)-f(x) \Big] \, dx 
\xrightarrow[j\into\infty]{}  z - \psi(P)  .
\end{align*}

\subsection{Variance}
%
functional $\psi_2(P)= P ( x-\psi_1(P))^2$ has influence function
$\wt\psi_2(x) = (x-\psi_1(P))^2 - \psi_2(P)$.

The policy sensitivity derivative of the mean with respect to the variance according to policy
metric $g$ as in \Cref{ex:mult} is
\begin{align*}
    \cp_{\psi_2}\psi_1(P) = \int 
    \Big[
	x - \psi_1 - P (x-\psi_1)\tfrac{dP}{dQ} / P\tfrac{dP}{dQ}
    \Big] \tfrac{dP}{dQ}(x) 
    \cdot 
    \big[
	(x - \psi_1)^2 -\psi_2
    \big]  \; dP(x).
\end{align*}

\subsection{\textit{p}-quantile}
%
of a continuous strictly increasing distribution is 
$\psi_3(P)=F_P^{-1}(p)$. IN formula allows to use paths through distributions with
these properties. Influence function can be derived from the following algebraic identity 
$ F_tF_t^{-1} (p) = p $ or
\begin{align*}
    (1-t)F\big( F_t^{-1}(p) \big) + t G_z^j \big( F_t^{-1}(p) \big) = p.
\end{align*}
Differentiating both sides with respect to $t$ and evaluating at $t=0$, obtain
\begin{align*}
    0 &= 
    \tfrac{d}{dt} F\big( F_t^{-1}(p) \big)_{\big|t=0} -F\big( F^{-1}(p) \big)
    + G_z^j \big( F^{-1}(p) \big) \\
    &=
    f\big( F^{-1}(p) \big) \cdot 
    \underbracket[0.2pt][2pt]{
    \tfrac{d}{dt} F_t^{-1}(p)_{\big|t=0} }_{ =\frac{d}{dt}\psi_3(P_t) }
    -F\big( F^{-1}(p) \big) + G_z^j \big( F^{-1}(p) \big).
\end{align*}
Solving for the $\frac{d}{dt}\psi_3(P_t)$, simplifying and taking limit on $j$, obtain
$\wt\psi_3(x)=\dfrac{p- 1_{[x,\infty)}(\psi_3(P))}{f(\psi_3(P))}$.
\vspace{1pt}

The policy derivative of the mean with respect to the $p$-quantile according to metric $g$ of
\Cref{ex:mult} is
\begin{align*}
    \cp_{\psi_3} \psi_1 
    =
    \frac{1}{f(\psi_3(P))} 
    \int
    \big[
	p - 1_{(-\infty, \psi_3]}(x)
    \big]
    \Big[
	x - \psi_1 -  P (x-\psi_1)\tfrac{dP}{dQ} / P\tfrac{dP}{dQ}   
    \Big] \tfrac{dP}{dQ}(x) \, dP(x)
    .
\end{align*}

\subsection{GMM.}
%
We study \textsc{gmm} functionals on the nonparametric model $\euP$ that is constrained only by
regularity (smoothness, integrability) conditions.
Application layer provides a parameter space
$\Theta\subset\mR^p$ and a vector of moment criterion functions
$$
g:\calX\times\Theta \into \mR^q.
$$

Specification layer maps the economic quantity $\vartheta\in\Theta$ to a function
$\psi:\euP\into\Theta$ of the statistical model. \textsc{gmm} estimation is setup from the
application layer assumptions that 
\begin{align}\label{eqn:moment_assumptions2}
    P g(\vartheta) = 0 \tag{$a_M$}.
\end{align}
This assumption is usually an optimality condition of the interactions described by the
application layer model. Often these models are highly stylized and are not expected to
describe real-world data precisely.
Our view is that this assumption should not be taken literally to data, and
that the role of specification layer is important and deserves attention
(but is beyond the scope of this paper).
We derive the sensitivity measures to provide a local characterization of a given
\textsc{gmm} functional.
Specifically we describe the local identification of \textsc{gmm} functionals
$\psi_W$ on the nonparametric model $\euP$ by measuring the local dependence of the estimated
parameter on the values of individual moments
$$
\nu_i(P) \coloneqq P g_i(\th)_{|\th=\psi_W}.
$$
The direction and absolute magnitude of the dependence is measured by the derivative
$\cp_{\nu(i)}\psi_W$.
The relative magnitude of dependence on $\nu_i$ to total local
variation in $\psi_W$ is measured by local sufficiency $R(\psi_W,\nu_i)$.  The latter also
measures the extent to which (statistical) uncertainty about the value of $\nu_i(P)$ in the model
$\euP$ determines inference about $\vartheta=\psi_W$ in the application layer.

Asymptotic distribution of misspecified \textsc{gmm} estimators was first considered tangentially
in \cite{imbens1997one} and derived explicitly in \cite{hall2003large}. We derive the influence
function (information gradient) of the functional
and use it to compute sensitivities. Our derivation provides a characterization of the tangent set
to the classical \textsc{gmm} model $\euP_0$ that is restricted by assumptions 
\eqref{eqn:moment_assumptions2} in the over-identified case $q>p$. As a bonus, this also shows
directly the semiparametric efficiency of `optimally weighted' estimator
$\wh\psi_{\Omega^{-1}}$ on $\euP_0$ and
of all \textsc{gmm} estimators $\psi_W$ of  \emph{different} functionals on the
nonparametric model $\euP$.
Although the values of functionals $\psi_W$ coincide on $\euP_0$ their
\emph{sensitivities} to directions ruled out by \eqref{eqn:moment_assumptions2} are
\emph{different}.
Chamberlain (1987) \cite{chamberlain1987asymptotic} first showed efficiency of over-identified
\textsc{gmm} estimators via discrete approximations.

We consider only deterministic weighting matrices $W$. In the over-identified case weighting
determines the functional and should be chosen based on application layer considerations (we call
this specification).  \textsc{gmm} functionals are defined by
$$
\psi_W(P) = \argmin_{\theta\in\Theta} P g(\theta)^T W Pg(\theta),
$$
or locally by the first order condition
\begin{align}\label{eqn:foc}
    \frac{\cp}{\cp\th} Pg(\th)^T W Pg(\th)_{\big|\th=\psi_W} = 0. \tag{\textsc{foc}}
\end{align}
To establish differentiability (relative to $H_2$ embedding) of the functional and to find the
influence function we assume it along with necessary regularity conditions to proceed with
a formal calculation that yields a candidate for the information gradient.
Once the gradient is found,
Riesz representation implies differentiability\footnote{this method has the name a priori estimate
in PDEs}.
Let $t\mapsto P_t$ be a smooth curve in $\euP$ with score vector $\xi\in L^2_0(P)$ at $t=0$. The
functional $\th_t=\psi_W(P_t)$ satisfies the \eqref{eqn:foc} along the curve $P_t$:
\begin{align}\label{eqn:foc2}
    P_t \Big[\frac{\cp}{\cp\th} g(\th_t) \Big]^T \, W \; P_t \Big[g(\th_t) \Big] = 0
    .
\end{align}
We use denominator layout for derivatives of vectors (so that $\cp g / \cp\th$ is $q$ by $p$); our
reference for matrix calculus is \cite{dhrymes1978mathematics}. Differentiating with
$\frac{d}{dt}$ in \eqref{eqn:foc2} obtain
%
%
\begin{align*}
    0 &=
    \frac{d}{dt} \Big[ P_t \frac{\cp}{\cp\th} g(\th_t)^T \, W \; P_t g(\th_t)
    \Big]_{\big|t=0}
    \\
    &=
    \Big( 
    \underbracket[0.2pt][2pt]{
	Pg(\th)^T W \otimes I_p 
    }_{\eqqcolon M}
    \Big) \;
    \underbracket[0.2pt][2pt]{
	\frac{d}{dt} \vect \Big[ 
	P_t \frac{\cp}{\cp\th} g(\th_t)^T
	\Big] 
    }_{I}
    +
    P \frac{\cp}{\cp\th} g(\th)^T \; W \;
    \underbracket[0.2pt][2pt]{
	\frac{d}{dt} \Big[ 
	P_t g(\th_t)
	\Big]
    }_{II}
    .
\end{align*}
Derivative $\frac{d}{dt}$ in terms $I,II$ has two components: perturbing distribution $P$ in the
direction $\xi$ changes the integrals and also the value of the functional $\psi_W$ which enters
the moment criterion functions.
%
%
Consider the first element of the vectorized term $I$ above
%
%
\begin{align*}
    I_1 &=
    \frac{d}{dt} \, \Big[ P_t \frac{\cp}{\cp\th_1} g_1(\th_t) \Big]_{\big|t=0}
    \\
    &=
    \int
    \underbracket[0.2pt][2pt]{
    \frac{\cp}{\cp\th} \Big(  \frac{\cp}{\cp\th_1} g_1(\th) \Big)
    }_{\;}
    \cdot \;\dot{\th} \; dP
    +
    \int   
    \underbracket[0.2pt][2pt]{
    \frac{\cp}{\cp\th_1} g_1(\th)
    }_{\;}
    \; \xi \; dP.
\end{align*}
Define the $qp\times p$ matrix $H_1$ and a $qp\times 1$ vector $H_2$ by stacking the underlined
terms in last screen
\begin{IEEEeqnarray*}{c"c}
    H_1 \coloneqq \frac{\cp}{\cp\th} \vect\Big[ \Big( \frac{\cp}{\cp\th} g(\th) \Big)^T \Big]
    &
    H_2 \coloneqq \vect \Big[ \Big( \frac{\cp}{\cp\th} g(x,\th) \Big)^T \Big]
\end{IEEEeqnarray*}
then
$$
I =  P[ H_1] \cdot \dot\th + P[H_2 \cdot \xi].
$$
Similarly
%
%
\begin{align*}
    II &=
    \frac{d}{dt} \Big[ \int g(\th_t) dP_t \Big]_{\big|t=0}
    = 
    \int \frac{\cp}{\cp\th} g(\th) \cdot \dot\th \; dP 
    +
    \int g(\th) \xi \; dP
    \\
    &=
    P[\frac{\cp}{\cp\th} g(\th) ] \cdot \dot\th 
    + 
    P[g(\th) \cdot \xi].
\end{align*}
Above manipulation implicitly assumes that $\th=\psi_W(P)$ is differentiable relative to the
embedding of statistical model into $H_2$. Recall that the differential $d\psi_W$ has a Riesz
representation
$$
\cp_\xi\th  = d\psi_W[\xi] = \brk{\wt\psi_W}{\xi}_{H_2} = \int \wt\psi_W \xi \; dP = \dot\th.
$$
The last equality is termed pathwise differentiability in bounds literature. The point of our work
in \cref{sec:geom} was to argue that the notion of differentiability used in bounds literature is
precisely the same as the one used with linear spaces and that directional
derivatives can be naturally interpreted.

By differentiating with $\frac{d}{dt}$ in \eqref{eqn:foc2} we obtained the following expression
that relates the pathwise (directional)  derivative $\cp_\xi\th$ and an integral involving the
tangent vector $\xi$:
\begin{IEEEeqnarray*}{c'c'l}
    0
    &=&
    \Big\{ 
	M \; P[H_1]
	+
	P[\cp_\th g(\th)^T] W P[\cp_\th g(\th)]
    \Big\} \cdot \dot\th
    +
    P\Big\{ 
	\Big( M  \; H_2 + P[\cp_\th g(\th)]^T W \; g(\th) \Big) \cdot \xi
    \Big\}
    .
\end{IEEEeqnarray*}
From above expression we can solve for the 
%
%
\begin{align}\label{eqn:GMM_grad_short}
    \wt\psi_W
    =
    - \Big[
	M \;
	H_1
	+
	P[\cp_\th g(\th)]^T W P[\cp_\th g(\th)]
    \Big]^{-1}
    \Big( 
	M \;
	H_2
	+
	P[\cp_\th g(\th)^T] W \; g(\th) 
    \Big).
\end{align}
Above derivation relies on smoothness and integrability conditions of moment functions $g$ and its
(parameter) derivatives. Since the tangent space $T_P\euP$ is unrestricted, we conclude that
\cref{eqn:GMM_grad_short} is the information gradient of $\psi_W$ and that functional is smooth
under these conditions.
Note that at $P\in\euP_0$ where moment assumptions \eqref{eqn:moment_assumptions2} hold, we have
$M=0$, which reduces the gradient to the familiar expression. Although at $P\in\euP_0$ all the
functionals $\psi_W$ obtained from different choices of weighting $W$ coincide, their gradients
are different along the directions $\xi$ that point outside the model $\euP_0$.

Assumptions \eqref{eqn:moment_assumptions2} imply restrictions for tangent set $T_P\euP_0$. We
characterize these restrictions next.
Differentiating along a path similarly to above in \eqref{eqn:moment_assumptions2}, obtain
$$
P g \, \xi = - G \cdot \dot\th,
\quad\text{where }
g\coloneqq g(\psi_w),
\quad
G\coloneqq P\cp_\th g(\th)_{\big|\th=\psi_W}
.
$$
This condition states that the change in the integral of criterion functions due to perturbing the
measure must be offset by the change in the value of the parameter. Since moments $Pg$ can move in
$q$ independent directions, where as parameter deviations $\dot\th$ can span only
$p=\text{rank}(G)$ of them, the condition is restrictive. Define continuous linear
operator
$$
    A: L^2_0(P) \into \mR^q
    \text{ by } A\xi \coloneqq Pg\xi,
    \quad \text{ so that }
    T_P\euP_0 = \Set{\xi\in L^2_0(P)}{A\xi \in R(G) }.
$$
We will derive projections $\Pi_0$ onto $T_P\euP_0 \subset L^2_0(P)$ and $\Pi_0^{\perp}$
onto the orthocomplement $T_P\euP_0(P)^\perp$.
First we reduce the problem to finite dimensional spaces by splitting
$$
L^2_0(P) = H_g \;\oplus\; H_g^\perp,
\quad \text{ where }
H_g \coloneqq \text{span} \set{g},
$$
and  noting that any vector $\xi\in L^2_0(P)$ that is orthogonal to $H_g$ does not change
the integral of the moment functions and therefore does not change the value of $\psi_W$,
as evident from \cref{eqn:GMM_grad_short}. Hence, $H_g^{\perp}\subset T_P\euP_0$.

It is then enough to consider operator $A:H_g\into \mR^q$ which is an isomorphism.
If $A\xi \in R(G)$ then $A\xi = G\th$ or $\xi = A^{-1}G\th$, therefore 
$$
H_g = H_G \oplus H_G^\perp
\quad \text{ where } \quad
H_G \coloneqq R(A^{-1}G), 
\quad
H_G^\perp \coloneqq N((A^{-1}G)^*)
$$
is the orthogonal decomposition of $H_g$ onto directions that are in $T_P\euP_0$ and those
that point outside the classical \textsc{gmm} model. We have the refined decomposition of
nonparametric tangent space:
$$
L^2_0(P) =
\underbracket[0.2pt][2pt]{
    H_g^\perp \oplus H_G
}_{T_P\euP_0}
\quad\oplus\quad
\underbracket[0.2pt][2pt]{
    \;\; H_G^\perp \;\;
}_{T_P\euP_0^\perp} .
$$

To compute the projection $\Pi_G$ onto the range $R((A^{-1}G)^*)$ we fix the orthonormal
basis $\Omega^{-1/2}g$, where $\Omega\coloneqq Pgg^T$,
then obtain the matrix of $A$ to be $A_{[\;]}=\Omega^{1/2}$ and apply
the regression formula for projection onto the range of $\Omega^{-1/2}G$
$$
\Pi_G = (\Omega^{-1/2}G) \big[ (\Omega^{-1/2}G)^T(\Omega^{-1/2}G) \big]^{-1}
(\Omega^{-1/2}G)^T,
\quad \text{ then }
\Pi_G^\perp = I_q - \Pi_G.
$$
Finally the projection $\Pi_0 \xi$ onto $T_P\euP_0$ of tangent vector $\xi\in L^2_0(P)$ is
obtained by removing the $H_G^\perp$ component that can be computed by passing to
coordinates and applying above projection matrix
$$
\Pi_0\xi = \xi - 
P[ \xi g^T\Omega^{-1/2} ] \;
\Big[ 
    I_q -
\Omega^{-1/2}G \big[ G^T\Omega^{-1}G \big]^{-1} 
G^T\Omega^{-1/2}
\Big] \;
\Omega^{-1/2} g.
$$
The classical \textsc{gmm} model restricts $q-p$ dimensions off of nonparametric tangent
space. Specifically, vectors of the form 
\begin{align}\label{eqn:gmm_out}
    \zeta = \a^T\Pi_G^\perp \Omega^{-1/2}g
\end{align}
are restricted, whose span is of dimension $\text{rank}(\Pi_G^\perp)$.

The efficient influence function for \textsc{gmm} on $\euP_0$ is obtained by
projecting any $\wt\psi_W$ in \cref{eqn:GMM_grad_short} onto the (mildly) restricted
$T_P\euP_0$:
\begin{align*}
    \Pi_0 \wt\psi_W 
    &=
    P \{(G^TWG)^{-1}G^TWg\} \;
    g^T\Omega^{-1/2} \;
    \Big[
    \Omega^{-1/2}G \big[ G^T\Omega^{-1}G \big]^{-1} G^T\Omega^{-1/2}
    \Big] \;
    \Omega^{-1/2} g
    \\
    &= (G^T\Omega^{-1}G)^{-1} G^T\Omega^{-1/2} \Omega^{-1/2} g
    = \wt\psi_{\Omega^{-1}}.
\end{align*}
Consequently, the sensitivity $\cp_\zeta \psi_{\Omega^{-1}}$ of the ``efficient''
\textsc{gmm} functional to any direction $\zeta$ that points out of the model $\euP_0$ is
zero, where as sensitivities of $\psi_W$ are nonzero. Estimators $\wh\psi_W$ suffer larger
asymptotic variance because they estimate the (local) values of the functional outside
of $\euP_0$ and have nonzero sensitivities to local deviations in those directions.

%
%
%


\section{Appendix} \label{sec:appendix}


In this section we collect results necessary to provide a self-contained proof of the convolution
theorem. My main two sources are \cite{vaart2000asymptotic,bickel1993efficient} but neither
provides an exposition that is both concise and self-contained.  For completeness I provide all
the details with minor variations on the proofs.

\subsection{Contiguity.}
%

Characterization of asymptotic distribution of estimators is achieved by requiring that the
convergence be sufficiently uniform. The limit distribution then is invariant under a sufficiently
rich class of converging sequences of probability measures.  These sequences provide complementary
pieces of information about the invariant limit distribution and allow for a sufficiently complete
characterization.
The property of sequences of probability measures that allows extracting information about the
limit distribution of a sufficiently robust estimator is an asymptotic counterpart of absolute
continuity. The idea is to be able to obtain limit distribution of estimator $T_n$ under sequence
of laws $Q_n$ from the limit distribution under laws $P_n$. 

Let $(\calX_n, \calA_n)$ be a sequence of sample spaces, we consider laws $Q_n$ and $P_n$ that are
dominated by sigma-finite measures $\mu_n$. Sequence $Q_n$ is \textbf{contiguous} to sequence
$P_n$, denoted $Q_n\lhd P_n$, if for every sequence of events $A_n$ with $P_n(A_n)\into 0$ it
holds that $Q_n(A_n)\into 0$. A good way to think about this definition is to interpret $A_n$ as critical
regions for testing $H_0: P_n$ against $H_1: Q_n$, then contiguity requires that there be no test
whose level gets close to zero and whose power stays bounded away from zero.

Let $Q^a_n\coloneqq \frac{dQ_n}{dP_n}dP_n$ and $Q^\bot_n\coloneqq Q_n - Q^a_n$ be the Lebesgue
decomposition of $Q_n$ with respect to $P_n$. The following proposition provides a low-tech
characterization of contiguity as asymptotic uniform absolute continuity that is intuitive and
useful in proofs.
\begin{proposition}
    \label{prop:contiguity_low_tech}
    The following are equivalent:
    \begin{enumerate}[label=(\roman*)]
	\item $Q_n \lhd P_n $;
	\item $Q^\bot_n(\calX) \into 0$ and $Q^a_n$ are uniformly absolutely continuous with
	    respect to $P_n$;
	\item $Q^\bot_n(\calX) \into 0$ and Radon-Nikodym derivatives $\frac{dQ_n}{dP_n}$ are
	    uniformly $P_n$-integrable;
    \end{enumerate}
\end{proposition}
\begin{pf}
    $(i)\Rightarrow(ii)$ From $P_n(\supp Q^\bot_n)=0$ have that $Q^\bot_n(\calX) \into 0$.
    Uniform absolute continuity means that for any $\e>0$ there is a $\d>0$ such that for any
    sequence of events $A_n$ it holds that $P_n(A_n)\le \delta$ implies $Q^a_n(A_n)\le\e$. This
    follows from $(i)$ by contradiction.

    Under $(ii)$, from Markov's inequality 
    \begin{align}
	\label{ineq:lratio_tightness}
	P_n\set{\tfrac{dQ_n}{dP_n}> M } \le \frac{P_n \, \frac{dQ_n}{dP_n}}{M} \le
	\frac{1}{M}
    \end{align}
    obtain uniform control on $P_n$ probabilities of the tail event and infer uniform bound on
    $Q^a_n\set{\frac{dQ_n}{dP_n}> M} \le \e$ for a suitable $M=M(\e)$. Uniform integrability in
    $(iii)$ follows immediately since
    \begin{align*}
	\int_{\set{\frac{dQ_n}{dP_n}> M}}  \frac{dQ_n}{dP_n} \, dP_n = Q^a_n
	\set{\tfrac{dQ_n}{dP_n}> M}.
    \end{align*}

    $(iii)\Rightarrow(i)$ Fix events $B_n$ with $P_n(B_n)\into 0$. Then
    \begin{align*}
	Q_n(B_n) &\le Q^\bot_n(\calX) + \int_{B_n} \frac{dQ_n}{dP_n} \, dP_n \\
	&\le Q^\bot_n(\calX) + \int_{B_n\cap \set{\frac{dQ_n}{dP_n} \le M}} \frac{dQ_n}{dP_n} \, dP_n
	+ \int_{\set{\frac{dQ_n}{dP_n}> M}}  \frac{dQ_n}{dP_n} \, dP_n \\
	&\le Q^\bot_n(\calX) + M P_n(B_n)
	    + \int_{\set{\frac{dQ_n}{dP_n}> M}}  \frac{dQ_n}{dP_n} \, dP_n \\
    \end{align*}
    can be made arbitrarily small by first choosing $M$ large enough to control the last term, and
    then demanding $n$ to be large enough to control the first two terms.
\end{pf}

Next we state a high-level characterization of contiguity that is useful in practice. Note that
the sequence of random variables $\frac{dQ_n}{dP_n}$ is tight under $P_n$ from
\cref{ineq:lratio_tightness}.
\begin{proposition}[(Le Cam, van der Vaart)]
    \label{prop:LeCam1}
    The following statements are equivalent:
    \begin{enumerate}[label=(\roman*)]
	\item $Q_n \lhd P_n $;
	\item If $\frac{dQ_n}{dP_n} \overset{P_n}{\rightsquigarrow} G$ along a subsequence, 
	    then $\int_\mR x \, dG =1;$
	\item If $\frac{dP_n}{dQ_n} \overset{Q_n}{\rightsquigarrow} F$ along a subsequence,
	    then $F\set{0} = 0$;
    \end{enumerate}
\end{proposition}
\begin{pf}

$(i) \Rightarrow (ii)$
Let $X_n,X_0$ be Skorohod representation of $\tfrac{dQ_n}{dP_n},G$. By
\cref{prop:contiguity_low_tech} $E X_n = Q^a(\calX)\into 1$; $X_n$ are uniformly integrable so
that $E X_n \into E X_0 =1 = \int_{\mR} x\;dG$.

$(ii) \Leftrightarrow (iii)$
Let $\mu_n = P_n + Q_n$, then along possibly further subsequences have limits
\begin{align*}
    W_n \coloneqq \tfrac{dP_n}{d\mu_n} \overset{\mu_n}{\rightsquigarrow} W,
    \tfrac{dQ_n}{dP_n} = \tfrac{1-W_n}{W_n} \overset{P_n}{\rightsquigarrow} G,
    \tfrac{dP_n}{dQ_n} = \tfrac{W_n}{1-W_n}\overset{Q_n}{\rightsquigarrow} F,
\end{align*}
Since $\tfrac{dP_n}{d\mu_n} \leq 1$ by bounded convergence have $1=\mu_n[ W_n ]\into
\int_{\mR} x \; dW$. For any $f\in\calC_b(\mR)$ the corresponding functions
$w\mapsto f(\tfrac{1-w}{w})w$ and $w\mapsto f(\tfrac{w}{1-w})(1-w)$ are also bounded and
continuoous on $[0,1]$. By assumed convergence in distribution
\begin{align*}
    \int f \; dG
    &=\lim_n E_{P_n} f(\tfrac{Q_n}{P_n})
    =\lim_n \int f(\tfrac{1-w}{w})w \; d\mu_n 
    =\int f(\tfrac{1-w}{w})w \; dW
    \shortintertext{and}
    \int f \; dF
    &=\lim_n E_{Q_n} f(\tfrac{P_n}{Q_n})
    =\lim_n \int f(\tfrac{w}{1-w})(1-w) \; d\mu_n
    =\int f(\tfrac{w}{1-w})(1-w) \; dW.
\end{align*}
By taking $0\le f_j \in \calC_b \uparrow x$ by monoton convergence obtain
\begin{align*}
    \int x \; dG =
    \int_{w>0} 1-w \; dW =
    W\set{w>0} - \int w \;  dW.
\end{align*}
  Similarly with $f_j \in \calC_b \downarrow 1_{x=0}$ by dominated convergence
\begin{align*}
    F\set{0} = W\set{w=0}.
\end{align*}
Therefore $\int x \;dG + F\set{0} =1$.

$(ii) \Rightarrow (i)$
Given $A_n$ with $P_n(A_n)\into 0$, choose critical regions $\phi_n = 1_{[\frac{dQ_n}{dP_n}>k_n]} +
\g_n1_{[\frac{dQ_n}{dP_n}=k_n]}$ with $P_n \phi_n = P_n(A_n)$ and $Q_n(A_n) \le Q_n\phi_n$. Then
for any $M>0$
\begin{align*}
    Q_n(A_n) \le Q_n \phi_n = \int_{[\frac{dQ_n}{dP_n}\le M]} \frac{dQ_n}{dP_n} \phi_n dP_n
    + \int_{[\frac{dQ_n}{dP_n}>M]} \phi_n dQ_n \\
    \le M \cdot P_n \phi_n + 
    1 - \int_{[\frac{dQ_n}{dP_n}\le M]} \frac{dQ_n}{dP_n} \; dP_n
\end{align*}
Arguing along a further convergent subsequence,  by bounded convergence 
\begin{align*}
    \int_{[\frac{dQ_n}{dP_n}\le M]} \tfrac{dQ_n}{dP_n} \; dP_n \into \int_{[x\le M]} x \; dG
\end{align*}
can be made arbitrarily close to $1$ by choice of large enough $M$ for all large $n$. Also
$M \cdot P_n \phi_n \into 0$. Conclude that $Q_n(A_n) \into 0$.
\end{pf}

We conclude with the result that contiguity was designed to provide: characterization of limit
distributions under contiguous deviations from the underlying sequence of probability measures.

\begin{proposition} \label{prop:LeCam3}
  If $Q_n\lhd P_n$ and $(X_n,\frac{dQ_n}{dP_n}) \overset{P_n}{\rightsquigarrow}(X,V)$, then
  $\int_\calX f(X_n) \, dQ_n \into  \mtt{E} f(X) V$ for every $f\in C_b(\calX)$.
\end{proposition}
\begin{pf}
By \cref{prop:LeCam1} and properties of Lebesgue integral 
\begin{align*}
  L(B) \coloneqq \mtt{E}[1_B(X) V]
\end{align*}
defines a probability measure. By monotone class theorem 
\begin{align*}
    \mtt{E}[f(X) V] = \int f(X) \; dL
\end{align*}
for every integrable function $f$.
By \cref{prop:contiguity_low_tech}, random variables $f(X_n)\frac{dQ_n}{dP_n}$ are
$P_n$-uniformly integrable and $Q_n^{\bot}(\calX)\into 0$ so that
\begin{align*}
  \int f(X_n) dQ_n 
  &= \int f(X_n) \frac{dQ_n}{dP_n} \; dP_n
  + \int f(X_n) \; dQ^{\bot}_n  \\
  &\into \mtt{E}[ f(X) V ] + 0 \\
  &= \int f(X) \; dL .
\end{align*}
Conclude that $X_n \overset{Q_n}{\rightsquigarrow} L$.
\end{pf}

\subsection{Regular parametric submodels.}
%
The differential structure on a statistical model $M$ that determines asymptotic distribution of
regular estimators is the one determined by imbedding $M$ into space $H_2$ of square roots of
measures. The variance bound for estimating $\psi(P)$ on $M$ is the operator norm of its
derivative. The bound is technically the supremum of the set of bounds for finite-dimensional
submodels. We consider smoothly parametrized finite-dimensional submodels and obtain convolution
representation on regular submodels.
The bound and convolution representation for the full semiparametric model $M$ is achieved on any
submodel that allows variation along gradient directions of the functional.

Let $(U,\bxi)$, where $U\subset \mR^m$ and $\bxi:U\into M$ be a local parametrization of a
submodel of $M$. Let $\mu$ be a dominating measure for the parametrized submodel. Define
\begin{align*}
  p_{\bxi} \coloneqq \frac{dP_{\bxi}}{d\mu} \quad\text{and}\quad
  s_{\bxi} \coloneqq 2\sqrt{p_{\bxi}}
  .
\end{align*}
Differentiability of root-density $s_{\bxi}=2\sqrt{p_{\bxi}}$ in $L^2(\mu)$ is defined in terms of
the norm, namely this requires existence of  measurable functions 
$\dot s_{\bxi} = (\dot s_{1,\bxi},\dots, \dot s_{m,\bxi}) \in L^2(\mu)$  that satisfy
\begin{align} \label{def:DQM}
  \int \Big[
    s_{\bxi + h} - s_{\bxi} - h^T \dot s_{\bxi}
  \Big]^2 \; d\mu
  = o(\abs{h}^2), \quad h\into 0.
\end{align}

\begin{definition} \label{def:reg_model}
  If above condition is satisfied, then the model is called differentiable in quadratic mean.  A
  statistical model that is a Riemannian manifold imbeddable  into $L^2(\mu)$ is called regular
  parametric.
\end{definition}

\begin{proposition}
  \label{DQM-properties}
  Model that is differentiable in quadratic mean has finite information matrix and
  $L^2(p_{\bxi} \mu)$ score functions that have zero mean.
\end{proposition}

\begin{pf}
  Information matrix elements 
  $I_{ij}(\bxi) \coloneqq \int \dot{s}_{i,\bxi} \dot{s}_{j, \bxi} \; d\mu$
  are finite by definition of DQM. Define 
  \begin{align*}
    \dot\ell_{\bxi} \coloneqq \frac{\dot s_{\bxi}}{\sqrt{p_{\bxi}}} 
    = 2\frac{\dot s_{\bxi}}{s_{\bxi}} 
  \end{align*}
  then $I_{ij}(\bxi) = \int \dot\ell_{i\bxi} \dot\ell_{j\bxi} \; d P_{\bxi}$, see
  \cite[A.5 prop 3]{bickel1993efficient}. Also DQM implies that
  $\sqrt{n}(s_{\bxi+h/\sqrt{n}} - s_{\bxi})$ converges to $h^T\dot s_{\bxi}$ in $L^2(\mu)$, and
  $s_{\bxi+h/\sqrt{n}}$ converges to $s_{\bxi}$. Then by continuity in $L^2(\mu)$
  \begin{align*}
    P_{\bxi} h^T \dot \ell_{\bxi}
    = \int h^T \dot s_{\bxi} \sqrt{p}_{\bxi} \; d\mu 
    = \lim \int \sqrt{n}(s_{\bxi+h/\sqrt{n}} - s_{\bxi}) \tfrac{1}{2}(s_{\bxi+h/\sqrt{n}} + s_{\bxi}) \; d\mu
  \end{align*}
  shows the score equality holds $P_{\bxi} \dot\ell_{\bxi} = 0 $.
\end{pf}

\subsection{Local asymptotic normality.}
%

A consequence of smoothness in parametric models is the validity of the following expansion
of likelihood ratios $dP^n_{\th+h/\sqrt{n}}/dP^n_{\th}$ of $n$-fold product measures at
distance $n^{-1/2}$ in local coordinates. Of primary interest to us here is the conclusion that
$P^n_{\th+h/\sqrt{n}}$ and $P^n_\th$ are mutually contiguous.


We adopt the following definition of likelihood ratios. Let $\mu=P+Q$, $p=\frac{dP}{d\mu}$ and
$q=\frac{dQ}d\mu$,
\begin{align*}
  \frac{dQ}{dP} \coloneqq \frac{p}{q} \mbf{1}_{\set{p>0}}
  + \mbf{1}_{\set{p=0}\cap\set{q=0}}
  + \infty \cdot \mbf{1}_{\set{q>0}\cap\set{p=0}} \in L^1(P).
\end{align*}

\begin{proposition} \label{prop:LAN}
  Let $M$ be a regular parametric model and $ \Theta\ni \th \mapsto \sb_\th \in L^2(\mu)$ be a local
  parametrization with derivative $\dot{\sb}_\th$. Then the following expansion holds
  \begin{align}
    \label{LAN-expansion}
    \log\frac{dP^n_{\th+h/\sqrt{n}}}{dP^n_\th} 
    = \frac{1}{\sqrt{n}}\sum_{i=1}^n h^T 2\frac{ \dot{\sb}_\th}{\sb_\th}
    - \tfrac{1}{2}h^T I_\th h + R_{n}(\th, h)
  \end{align}
  where the remainder term satisfies $R_n(\th,h)\xrightarrow{P^n_\th} 0$  uniformly for
  $h\in K\subset\subset \mR^m$, and if $\dot{\sb}_\th$ is continuous, then also uniformly for $\th\in
  K\subset\subset\Theta$.
\end{proposition}

\begin{pf}
  Proof is based on Taylor's series with Lagrange's remainder of third order.
  Uniformity of convergence for $\th$ on compacts is a consequence of compactness in $L^2(\mu)$.

  Define the following random variables and events on the product sample space
  \begin{align*}
  W_{ni}(\th,h) &\coloneqq 2\big(\frac{\sb_{\th+h/\sqrt{n}}}{\sb_\th}(x_i) - 1\big) \in
  L^2(P_\th)  \\
  A_{n}(\th,h) &\coloneqq \set{\max_{1\le i\le n} \abs{W_{ni}(\th,h)} \le \e} 
  \end{align*}
  In part $(i)$ we show that $P^n_\th(A_n^\complement) \xrightarrow[n\into\infty]{} 0$ with uniformity
  according to smoothness of $M$, therefore it suffices to prove $\cref{LAN-expansion}$ on
  events $A_n$, where we expand $\log (1+x)=x-\frac{x^2}{2}+ \frac{1}{3(1+\xi)^3}x^3$ with
  $\xi$ between $0$ and $x$:
  \begin{align*}
      \log\frac{dP^n_{\th+h/\sqrt{n}}}{dP^n_\th}
      &= \sum_{i=1}^n 2\log(1+\tfrac{1}{2}W_{ni}) \\
      &= \underbracket[0.2pt][2pt]{ \sum_{i=1}^n W_{ni} }_{\text{ part }(iv)\;}
      -  \underbracket[0.2pt][2pt]{ \tfrac{1}{4} \sum_{i=1}^n W^2_{ni} }_{\text{part }(ii)}
      +  \underbracket[0.2pt][2pt]{ \tfrac{1}{4} \sum_{i=1}^n \a_{ni} W^3_{ni} }_{\text{part
      }(iii)} .
  \end{align*}
  Part (i). We claim that
  \begin{align*}
    \sup_{h\in K\ssubset\mR^N} P^n_\th(A_n(\th,h)^\complement) 
    &\xrightarrow[n\into\infty]{} 0
    \quad \text{if $M$ is regular} \\
    \sup_{h\in K\ssubset\mR^N} \sup_{\th\in K\ssubset\Theta} P^n_\th(A_n(\th,h)^\complement)
    &\xrightarrow[n\into\infty]{} 0
    \quad \text{if $M$ has continuous tangent planes}. 
  \end{align*} 
  This follows from
  \begin{align*}
    P_\th \set*{\abs{W_{ni}} > \e} 
    &\le P_\th\set*{ \abs{W_{ni}- 2\frac{\dot\sb_\th\frac{h}{\sqrt{n}}}{\sb_\th}} > \tfrac{\e}{2} }
    + P_\th \set*{ \abs{2\frac{\dot\sb_\th\frac{h}{\sqrt{n}}}{\sb_\th}} > \tfrac{\e}{2} } \\
    &\le \frac{ \norm{\sb_{\th+h/\sqrt{n}} - \sb_\th - \dot\sb_\th
      \tfrac{h}{\sqrt{n}} }^2_{L^2(\mu)} }{\e^2/16}
    + \tfrac{\abs{h}^2}{\e^2 n}  \int_\calX  \abs*{ \dot\sb_\th }^2
    \mbf{1}_{ \set*{ \abs*{ \dot{\sb_\th}} > \tfrac{\e\sqrt{n}}{4\abs{h}} } } \,d\mu.
  \end{align*}
  The first term is of order $o(\frac{\abs{h}^2}{n})$, uniformly over compacts in $\th$ under
  continuous differentiability.
  For the second term we note that $\Set{\dot{\sb}_\th}{\th\in K\ssubset \Theta}$ is a compact
  subset of $L^2(\mu)$ under continuous differentiability and therefore
  uniformly integrable \cite{van2014compactness}. Claims follow by a union bound with $n$ terms.

  Part (ii). Here everything converges in $L(P_\th)$ norm.
  \begin{align*}
      \sum_{i=1}^n W_{ni}^2
      &= \sum_{i=1}^n \Big( W_{ni} - \tfrac{1}{\sqrt{n}} h^T 2\frac{\dot\sb_\th}{\sb_\th} 
	  + \tfrac{1}{\sqrt{n}} h^T 2\frac{\dot\sb_\th}{\sb_\th} \Big)^2  \\
      &= \sum_{i=1}^n \Bigg[ 
	\underbracket[0.2pt][2pt]{ 
	  \Big( \frac{ \sb_{\th+h/\sqrt{n}} -  \sb_\th - \tfrac{1}{\sqrt{n}} h^T \dot\sb_\th
	       }{ \sb_\th } \Big)^2 
	      }_{\text{term } A}
	  +2 \underbracket[0.2pt][2pt]{
	  \Big(  \frac{ \sb_{\th+h/\sqrt{n}} -  \sb_\th - \tfrac{1}{\sqrt{n}} h^T \dot\sb_\th
	     }{ \sb_\th } \Big) 
	  \Big( \frac{ \tfrac{1}{\sqrt{n}} h^T \dot\sb_\th}{\sb_\th} \Big)
	      }_{\text{term } C}
	  + \underbracket[0.2pt][2pt]{
	    \Big(  \frac{ \tfrac{1}{\sqrt{n}} h^T \dot\sb_\th}{\sb_\th} \Big)^2
	      }_{\text{term } B}
	  \Bigg] \\
      &\into   h^T I_\th h.
  \end{align*}
  Term $A$ is of order $o(\frac{\abs{h}^2}{n})$ by differentiability in $L^2(\mu)$.
Term $B$ converges by LLN: 
$\frac{1}{n}\sum_i \Big(\frac{ h^T \dot\sb_\th}{\sb_\th} \Big)^2 \overset{L^1(P_\th)}{\into} P_\th
(h^T \dot\sb_\th)^2 = h^T I_\th h $.
Term $C$ is of order $o(\tfrac{\abs{h}^2}{n}) O(1)$ by
Cauchy-Schwarz.

  Part (iii).
  This part is controlled in probability.
  \begin{align*}
    \abs*{\sum_{i=1}^n \a_{ni} W^3_{ni} }
    \le \max_{1\le i \le n} \abs{\a_{ni} W_{ni}} \cdot \sum_{i=1}^n W_{ni}^2
    = o_{P_\th}(1) O(1)
  \end{align*}
  by part(i) and part(ii).

  Part (vi). This is the main term, recall
  $ W_{ni}(\th,h) = 2\big(\frac{ \sb_{\th+h/\sqrt{n}} - \sb_\th }{ \sb_\th } \big) 
  \sim 2 \frac{ \tfrac{1}{\sqrt{n}} h^T \dot\sb_\th }{ \sb_\th } $. Let's compare their first
  moments:
  \begin{align*}
    P^n_\th \Big( \sum  2 \tfrac{1}{\sqrt{n}} h^T\frac{ \dot\sb_\th  }{ \sb_\th } \Big)
    &= 2n \int \sb_\th \; \tfrac{1}{\sqrt{n}}  h^T \dot\sb_\th \, d\mu  = 0 \\
    P^n_\th \Big( \sum 2\frac{ \sb_{\th+h/\sqrt{n}} - \sb_\th }{ \sb_\th } \Big)
    &= 2n\int (\sb_{\th+h/\sqrt{n}} - \sb_\th)\sb_\th \, d\mu  \\
    &= -n \int ( \sb_\th^2  -2\sb_\th \sb_{\th+h/\sqrt{n}} 
      + \sb_{\th+h/\sqrt{n}}^2 ) \, d\mu \\
    &= -n \norm{ \sb_{\th+h/\sqrt{n}} - \sb_\th }^2  \\
    &\into  - \norm{ h^T \dot\sb_\th }^2 = -\tfrac{1}{4}h^T I_\th h.
  \end{align*}
  We expect these sums to get close after removing the difference in means:
  \begin{align*}
      P^n_\th \Bigg[ 
	  \sum  2\frac{ \tfrac{1}{\sqrt{n}} h^T \dot\sb_\th }{ \sb_\th } 
	  - \Big( 2\sum \frac{ \sb_{\th+h/\sqrt{n}} - \sb_\th }{ \sb_\th }
      - \tfrac{1}{4}h^T I_\th h \Big)
      \Bigg]^2
      &= \mrm{Var}_\th(--) + \Big[ \mathtt{E}_\th(--) \Big]^2 \\
      &= n \mrm{Var}_\th(
	  \frac{ \sb_{\th+h/\sqrt{n}} - \sb_\th }{ \sb_\th }
	  -
	  \frac{ \tfrac{1}{\sqrt{n}} h^T \dot\sb_\th }{ \sb_\th } 
	  ) + \Big[ \mathtt{E}_\th(--) \Big]^2 \\
	  &\le n \norm*{ \sb_{\th+h/\sqrt{n}} - \sb_\th - h^T \dot\sb_\th  }^2_{L^2(\mu)} + o(1)
	  =o(1).
  \end{align*}
  by above analysis of the expectation term and differentiability in $L^2(\mu)$ hypothesis.
  Conclude 
  \begin{align*}
      \sum W_{ni} - 
      \sum  \tfrac{1}{\sqrt{n}} h^T  2\frac{ \dot\sb_\th }{ \sb_\th }
	  + \tfrac{1}{4}h^T I_\th h
	  \overset{L(P_{\th})}{\into} 0.
  \end{align*}
\end{pf}

From LAN expansion (\ref{LAN-expansion}) we see that likelihood ratios 
$$
\log dP^N_{\th+h/\sqrt{n}} / dP^n_{\th} \overset{P^n_{\th}}{\rsa} N(-\tfrac{1}{2} h^T I_{\th} h^T, h^T
I_\th h^T)
$$
converge in distribution, therefore by \ref{prop:LeCam1} sequences of laws $P^n_{\th+h_n/\sqrt{n}}$ and
$P^n_{\th}$ are mutually contiguous.

\subsection{Convolution theorem.} \label{convolution2}
%

An estimator $T_n$ of a functional $\varphi:M\into \mR$ is regular at $P\in M$ if 
\begin{align*}
  \sqrt{n}\big( T_n - \varphi_{\bxi(n)} \big) \overset{P^n_{\bxi(n)}}{\rsa} L_{\bxi}
\end{align*}
whenever $\sqrt{n}(\bxi_n -\bxi) = O(1)$.
Regularity on a semiparametric model is just regularity on every regular submodel.
This is a uniformity requirement, similar to uniform
unbiasedness condition of CR-bound. In particular
\begin{align}
    \sqrt{n}\big( T_n - \varphi_{\bxi} \big) 
    = \sqrt{n} \big( T_n - \varphi_{ \bxi + h/\sqrt{n} } \big)
    +
    \tfrac{1}{n^{-1/2}} \big(\varphi_{\bxi + hn^{-1/2}} - \varphi_{\bxi} \big)
    \overset{P^n_{\bxi(n)}}{\rsa} L_{\bxi} \ast \d_{ h^T \dot{\varphi} }.
    \label{regularity}
\end{align}
Thus regularity is an asymptotic condition of local unbiased.
Using samples from the perturbed sequence of laws with a regular estimator has the effect of
shifting the asymptotic distribution of estimates linearly in the direction of perturbation
according to the derivative of the target functional.
In the limit, perturbation is on the tangent plane $T_{P(\bxi)} M$ in the direction 
$h^T \dot{s}_{\bxi}$,
and changes the value of the functional by
$d\varphi_P  \big(h^T\dot{s}_{\bxi} \big) = g_P(\nabla \varphi, h^T\dot{s}_{\bxi})$.
A regular estimator is required to honestly reflect such deviation by recentering its
asymptotic distribution around the new value.
The following theorem provides an asymptotic version of a lower bound on efficiency of regular
estimators and a connection between geometry of statistical models and inference.

\begin{thmSc} \label{convolution}
Let $T_n$ be a regular estimator of a smooth functional $\varphi:M\into \mR^d$ on a regular parametric
model. Then
\begin{gather}
    \label{jointConv}
    \begin{pmatrix} 
	\sqrt{n}\big(T_n - \varphi_{\bxi} \big) - 
	\tfrac{1}{\sqrt{n}} \sum_{i=1}^n \nabla \varphi_{\bxi} \\
        \tfrac{1}{\sqrt{n}}\sum_{i=1}^n \nabla \varphi_{\bxi}
    \end{pmatrix}  
    \overset{P^n_{\bxi}}{\rightsquigarrow }
    \Delta_{T,\bxi} \times
    N\big(0,  (\cp_{\varphi_j} \varphi_i)_{ij} \big)
\shortintertext{so that}
    \label{Conv}
    \sqrt{n}\big(T_n - \varphi_{\bxi} \big)
\overset{P^n_{\bxi}}{\rightsquigarrow } L_{\bxi} =
    N\big(0,  (\cp_{\varphi_j} \varphi_i)_{ij} \big) \ast \Delta_{T,\bxi}.
\end{gather}
Here $\nabla\varphi$ denotes the vector of  gradients
$\big( \dot{\varphi}_{i,\bxi}^T I^{-1}_{\bxi} \dot \ell_{\bxi} \big)_i$
of target functionals expressed in local coordinates \eqref{eq:gradFormula}; and 
$\big(\cp_{\varphi_j} \varphi_i \big)_{ij} =\big( g_p(\nabla\varphi_i,\nabla\varphi_j)\big)_{ij}$
denotes the matrix of directional derivatives of the target functionals with respect to each
others gradient directions \eqref{eqn:cpInfluenceFcts}.
\end{thmSc}

\begin{proof}
Follows closely \textcite[p24-26]{bickel1993efficient}.
Let $(U_n, V_n) = \big(\sqrt{n}(T_n - \varphi_{\bxi}), n^{1/2}\sum_{i=1}^n
\ell_{\bxi}(X_i)\big)$. By assumed regularity of the estimator and the model, the sequence
marginally convergence in distribution. By Prohorov's theorem the sequence is marginally tight
and therefore jointly tight by a union bound. By examining an arbitrarily subsequential limit
$(U,V)$
and showing that it is unique we will conclude that the whole sequence converges in
distribution under $P^n_{\bxi}$. Here $U \sim L_{\bxi}$ and $V\sim N(0,I_{\bxi})$ but the
joint distribution possibly depends on the subsequence.

By LAN of the model \cref{prop:LAN}
\begin{align*}
    W_n \coloneqq \ell_n(\bxi+h/\sqrt{n}) - \ell_n(\bxi) &=
    \tfrac{1}{\sqrt{n}} \sum_{n} h^T \dot{\ell}_{\bxi}(X_i) -\tfrac{1}{2} h^T I_{\bxi} h + o_{P}(1)
    \\
    &= h^T V - \tfrac{1}{2} h^T I_{\bxi} h + o_{P}(1).
\end{align*}
Therefore by continuous mapping $(U_n,e^W_n)\rightsquigarrow (U,e^{h^T V - \tfrac{1}{2} h^T
I_{\bxi} h})$ which shows contiguity $P^n_{\bxi} \lhd \rhd P^n_{\bxi + h/\sqrt{n}}$ by
\cref{prop:LeCam1}.
Next we use regularity of the estimator together with contiguity to characterize the
subsequential joint limit. By regularity:
\begin{align}
    \sqrt{n} (T_n - \varphi_{\bxi + h/\sqrt{n}} )  \overset{P^n_{\bxi+h/\sqrt{n}}}{\rightsquigarrow} 
    L_{\bxi} \ast \d_{\dot{\varphi} h}, \nonumber
    \shortintertext{so that by Portmanteau}
    \label{jointCHF1}
    P_{\bxi + h/\sqrt{n}} [ e^{i a^T U_n}] \into  
    E [ e^{i a^T U} \cdot e^{i a^T \dot{\varphi} h} ].
\end{align}
Now by contiguity, [LeCam] and Portmanteau we compute the limit under alternative to be
\begin{align}
    \label{jointCHF2}
    P_{\bxi + h/\sqrt{n}} [ e^{i a^T U_n}] \into  
    E [e^{i a^T U} e^{h^T V - \tfrac{1}{2}h^T I_{\bxi} h}].
\end{align}
So regularity of the model (via LAN) and regularity of the estimator (via contiguity) provide
complementary characterizations of the joint characteristic function of $(U,V)$.
The limit in \eqref{jointCHF1} is a
holomorphic function several complex variables $h\in \mC^m$ for any fixed $a\in\mR^d$. The limit
in \eqref{jointCHF2} is a uniformly convergent over compact sets weighted average (gaussian
integral) of holomorphic
functions of $h\in\mC^m$ for any fixed $a\in\mR^d$. By analytic continuation off of $h\in\mR^m$
conclude that the two limits agree on $\mC^m$.
%
For $h=-iI^{-1}_{\bxi}\dot{\varphi}^T(a-b)$ we obtain the following expression for the joint
characteristic function of the limit distribution in \eqref{jointConv}
\begin{align}
    \label{jointCHF3}
    E\Big[ e^{i a^T(U - \dot{\varphi}I^{-1}_{\bxi} V) + ib^T \dot{\varphi} I^{-1}_{\bxi} V}
    \Big] =
    E \Big[ 
	e^{ia^T U}
    \Big]
	e^{ \tfrac{1}{2} a^T\dot\varphi I^{-1}_{\bxi} \dot\varphi a}
	e^{ -\tfrac{1}{2} b^T\dot\varphi I^{-1}_{\bxi} \dot\varphi b}
	\quad a,b\in\mR^d.
\end{align}
Since the subsequential limit distribution in \eqref{jointConv} is unique, conclude that the
entire sequence converges with the limit given in last screen. By setting $b=0$ we obtain the
characteristic function of $U - \dot{\varphi}I^{-1}_{\bxi} V$
\begin{align}
    \label{marginalCHF1}
    E\Big[ e^{i a^T(U - \dot{\varphi}I^{-1}_{\bxi} V) }
    \Big] =
    E \Big[ 
	e^{ia^T U}
	e^{ \tfrac{1}{2} a^T\dot\varphi I^{-1}_{\bxi} \dot\varphi a}
    \Big]
	\quad a\in\mR^d.
	\shortintertext{similarly with $a=0$, the characteristic function of
	$\dot\varphi_{\bxi} I^{-1}_{\bxi}V$ is}
    \label{marginalCHF2}
    E\Big[ e^{ib^T \dot{\varphi} I^{-1}_{\bxi} V}
    \Big] =
    E \Big[ 
	e^{ -\tfrac{1}{2} b^T\dot\varphi I^{-1}_{\bxi} \dot\varphi b}
    \Big]
	\quad b\in\mR^d.
\end{align}
Combining \eqref{jointCHF3},\eqref{marginalCHF1} and \eqref{marginalCHF2}, conclude that 
$U - \dot{\varphi}I^{-1}_{\bxi} V$ and $ \dot\varphi_{\bxi} I^{-1}_{\bxi}V$ are independent
according to the limit law in \eqref{jointConv}. Also since \eqref{marginalCHF2} is the
characteristic function of a 
$N\big(0, (\cp_{\varphi_j} \varphi_i)_{ij} \big)$ conclude representation \eqref{jointConv}.
\end{proof}

\end{spacing}

\begin{spacing}{0.9}
\renewcommand*{\bibfont}{\small}
\nocite{*}
\printbibliography

\end{spacing}

\thispagestyle{articleStyle}

\end{document}